\documentclass[fleqn,usenatbib]{mnras}

\usepackage{newtxtext,newtxmath}
\usepackage{subcaption}
\usepackage{graphicx}
\usepackage{multirow}
\usepackage{soul}
\usepackage[T1]{fontenc}

\DeclareRobustCommand{\VAN}[3]{#2}
\let\VANthebibliography\thebibliography
\def\thebibliography{\DeclareRobustCommand{\VAN}[3]{##3}\VANthebibliography}

\usepackage{graphicx}	% Including figure files
\usepackage{amsmath}	% Advanced maths commands
\usepackage{adjustbox}  % For the orcid ids
\makeatletter % Workaround to only color the year for cite commands
  \patchcmd{\NAT@citex}
    {\@citea\NAT@hyper@{%
      \NAT@nmfmt{\NAT@nm}%
      \hyper@natlinkbreak{\NAT@aysep\NAT@spacechar}{\@citeb\@extra@b@citeb}%
      \NAT@date}}
    {\@citea\NAT@nmfmt{\NAT@nm}%
    \NAT@aysep\NAT@spacechar\NAT@hyper@{\NAT@date}}{}{}

  \patchcmd{\NAT@citex}
    {\@citea\NAT@hyper@{%
      \NAT@nmfmt{\NAT@nm}%
      \hyper@natlinkbreak{\NAT@spacechar\NAT@@open\if*#1*\else#1\NAT@spacechar\fi}%
        {\@citeb\@extra@b@citeb}%
      \NAT@date}}
    {\@citea\NAT@nmfmt{\NAT@nm}%
    \NAT@spacechar\NAT@@open\if*#1*\else#1\NAT@spacechar\fi\NAT@hyper@{\NAT@date}}
    {}{}
\makeatother

\newcommand\orcid[1]{\href{http://orcid.org/#1}{\adjustbox{trim={-.15\width} {0\height} {-.15\width} {0\height},clip}{\includegraphics[height=12pt]{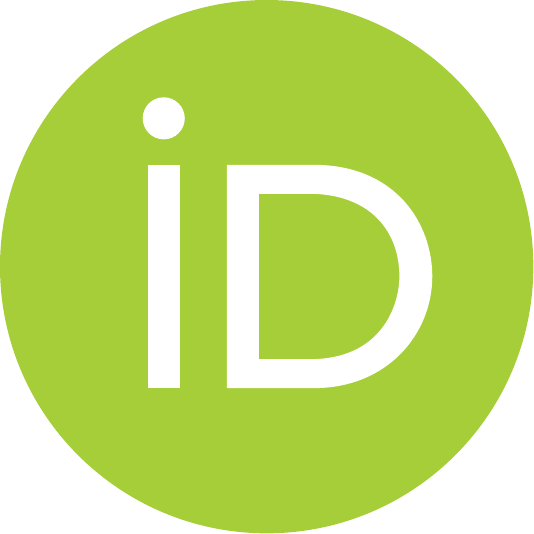}}}}

\usepackage[pagewise]{lineno}
\title[\ion{O}{i} absorbers in \textsc{thesan-zoom}]{The \textsc{thesan-zoom} project: The Hidden Neighbours of \ion{O}{i} Absorbers during Reionization}

\author[G. Pruto et al.]{%
Giulia~Pruto\orcid{0000-0003-0828-7192},$^{1}$\thanks{E-mail: \href{mailto:giulia.pruto@ed.ac.uk}{giulia.pruto@ed.ac.uk}}
Laura~Keating\orcid{0000-0001-5211-1958},$^{1}$
Rahul~Kannan\orcid{0000-0001-6092-2187},$^{2}$
Ewald~Puchwein\orcid{0000-0001-8778-7587},$^{3}$
Aaron~Smith\orcid{0000-0002-2838-9033},$^{4}$
\newauthor
Josh~Borrow\orcid{0000-0002-1327-1921},$^{5}$
Enrico~Garaldi\orcid{0000-0002-6021-7020},$^{6}$
Mark~Vogelsberger\orcid{0000-0001-8593-7692},$^{7}$
Oliver~Zier\orcid{0000-0003-1811-8915},$^{8,7}$
\newauthor
William~McClymont\orcid{0009-0009-5565-3790},$^{9, 10}$
Xuejian~Shen\orcid{0000-0002-6196-823X}$^{7}$
and Sandro~Tacchella\orcid{0000-0002-8224-4505}$^{9, 10}$
\\
\\
$^{1}$Institute for Astronomy, University of Edinburgh, Blackford Hill, Edinburgh, EH9 3HJ, UK\\%
$^{2}$ Department of Physics and Astronomy, York University, 4700 Keele Street, Toronto, ON M3J 1P3, Canada\\%
$^{3}$ Leibniz-Institut f\"ur Astrophysik Potsdam, An der Sternwarte 16, 14482 Potsdam, Germany\\%
$^{4}$ Department of Physics, The University of Texas at Dallas, Richardson, Texas 75080, USA\\%
$^{5}$ Department of Physics and Astronomy, University of Pennsylvania, 209 South 33rd Street, Philadelphia, PA 19104, USA\\%
$^{6}$ Kavli Institute for the Physics and Mathematics of the Universe, The University of Tokyo, 5-1-5 Kashiwanoha, Kashiwa, 277-8583, Chiba, Japan\\%
$^{7}$ Department of Physics, Kavli Institute for Astrophysics and Space Research, Massachusetts Institute of Technology, Cambridge, MA 02139, USA\\%
$^{8}$ Center for Astrophysics | Harvard \& Smithsonian, 60 Garden St, Cambridge, MA 02138, USA\\%
$^{9}$ Kavli Institute for Cosmology, University of Cambridge, Madingley Road, Cambridge CB3 0HA, UK\\%
$^{10}$ Cavendish Laboratory, University of Cambridge, 19 JJ Thomson Avenue, Cambridge, CB3 0HE, UK%
}

\date{Accepted XXX. Received YYY; in original form ZZZ}

\pubyear{\the\year{}}

\begin{document}
\label{firstpage}
\pagerange{\pageref{firstpage}--\pageref{lastpage}}
\maketitle

\begin{abstract}
Metal absorbers represent a powerful probe of galaxy feedback and reionization, as highlighted by both observational and theoretical results showing an increased abundance of low-ionised metal species at higher redshifts. 
The origin of such absorbers is currently largely unknown because of the low number of galaxy counterparts detected, suggesting that they might be surrounded by low-mass faint sources below the current detection threshold.
We use the \textsc{thesan-zoom} radiation hydrodynamic simulations to investigate the connection between properties of neutral oxygen (\ion{O}{i}) absorbers and galaxies across redshift $z = 5 - 8$. 
We find that the virialised gas in haloes becomes progressively ionised with cosmic time, leading to a decrease of $\approx 0.2$ in the covering fraction of neutral oxygen, while the total oxygen covering fraction remains constant.
Comparing the \ion{O}{i} line density obtained from our covering fractions with the trend suggested by blind quasar observations, we determine that the observable absorbers ($N_{\ion{O}{I}} \gtrsim 10^{13}\,\text{cm}^{-2}$) are not confined to haloes: at $z \geq 5$ the majority ($\gtrsim 60\%$) arise beyond $R_{\mathrm{vir}}$, 
consistent with recent JWST results.
Close to \ion{O}{i} absorbers, low-mass galaxies ($M_\star \leq 10^8\,\rm{M}_\odot$) are more commonly found, while, depending on the simulated environment, we do not exclude the possibility of nearby more massive star-forming sources ($\geq 5\,\text{M}_\odot\,\text{yr}^{-1}$) similar to those suggested by the latest ALMA observations. 
These results establish \ion{O}{i} absorbers as sensitive tracers of the evolving ionisation structure around faint galaxies to be probed by forthcoming deep spectroscopic surveys.
\end{abstract}


%%%%%%%%%%%%%%%%%%%%%%%%%%%%%%%%%%%%%%%%%%%%%%%%%%

%%%%%%%%%%%%%%%%% BODY OF PAPER %%%%%%%%%%%%%%%%%%
\section{Introduction}
\label{sec:Introduction}

The Epoch of Reionization (EoR) is the period during which the first stars and galaxies lit up the Universe, releasing high-energy radiation that transformed the intergalactic medium (IGM) from cold and neutral to warm and ionized. 
Understanding the history and topology of reionization is essential to gain insight into the formation and evolution of the earliest sources \citep{Wise2019, Robertson2022}.
Traditionally, constraints come from measurements of the neutral hydrogen (\ion{H}{I}) fraction, through the analysis of Ly$\alpha$ in absorption \citep[][]{GunnPeterson1965} and in emission \citep[][]{McQuinn2007}.
Recently, more sophisticated techniques involving the investigation of Ly$\alpha$ and Ly$\beta$ dark pixels and dark gaps \citep[][]{Zhu2022, Jin2023}, together with the study of transmission spikes \citep[][]{Garaldi2019, Gaikwad2020, Yang2020} have revealed the patchy nature of reionization.
Further efforts also highlighted the presence of islands of neutral gas down to $z\sim5.3$, when reionization is considered mostly complete \citep[][]{Bosman2022, Zhu2023}.
At higher redshift, above $z\approx6.3$, where the Ly$\alpha$ forest saturates, constraints rely instead on damping wing analyses \citep[][]{MiraldaE1998, Wang2020, Greig2022, Greig2024} and on observations of Ly$\alpha$ emitters \citep[LAEs;][]{Konno2014, Konno2018, Itoh2018, Ouchi2020, Kageura2025}.
Results from these measurements favour a late and rapid reionization scenario, but they remain subject to large uncertainties, making it challenging to establish a robust theoretical model.

Complementary information comes from the study of the ionisation state of metals in the IGM and, closer to the galaxies, in the circumgalactic medium (CGM), i.e. the gas gravitationally bound to haloes. 
In fact, the formation of the first sources not only produced intense UV radiation, causing the onset of reionization, but early stellar feedback and supernovae also drove winds able to enrich the surrounding gas with metals.
Although at high redshift we would expect to probe metal-poor gas due to the fact that there has been less time for star formation activity to enrich the surroundings, observations revealed how the line density of low-ionised species such as \ion{O}{I}, \ion{C}{ii}, \ion{Si}{ii}, \ion{Mg}{ii} remains constant or slightly rises with increasing redshift \citep[][]{Becker2011, Becker2019, Sebastian2024}.
In contrast, highly-ionised metals like \ion{C}{iv} and \ion{Si}{iv} display a substantial decrease \citep[][]{Ryan-Weber2009, Simcoe2011, Cooper2019, D'Odorico2022, Christensen2023}.
These pieces of evidence suggest a major change in the ionisation state of metals, with low-ionisation species preferentially found at high redshift, showing how the imprint of reionization can be traced through the line density evolution of heavy elements \citep[][]{Becker2015, Fan2023}.

Yet, finding the galaxies that produced and ionized these elements represents a significant observational challenge.
In fact, multiple factors such as the star-formation rate (SFR), outflow velocity, and mass-loading factors contribute to the metal enrichment of the CGM, and while theoretical models predict most absorbers at high redshift to reside nearby the more numerous low-mass galaxies \citep[e.g.,][]{Keating2014, Garcia2017}, detecting such faint sources in observations can be very demanding, as it would require deep and long exposures.
Ground-based narrow-band imaging and spectroscopy highlight the presence of low-luminosity Ly$\alpha$ emitters around strong \ion{C}{iv} absorbers at $z = 5.7$ \citep[][]{Diaz2014, Diaz2015, Diaz2021, Cai2017}, suggesting that highly-ionised absorbers might reside close to young and low-mass galaxies.
Higher redshift windows ($z\gtrsim6$) are now increasingly accessible thanks to the \textit{James Webb Space Telescope} (\textit{JWST}), and so far only medium and high-mass galaxies with impact parameters ranging from $25$ to $\gtrsim300$\,kpc from absorbers have been identified at $z\approx6$ \citep[][]{Zou2024, Higginson2025}. 
Moreover, studies show how most of the \ion{O}{i} and \ion{Mg}{ii} absorbers during reionization are not bound to detected sources \citep[][]{Bordoloi2023}, with more than half of the detectable neutral oxygen absorbers expected to reside close to low-mass undetected galaxies or in dense neutral IGM pockets \citep[][]{Higginson2025}.
Complementary results have come from the Atacama Large Millimeter Array (ALMA), which has uncovered [\ion{C}{ii}]-emitting galaxies in some quasar fields of view \citep[e.g.,][]{Kashino2023nat}. Notably, one of those sources lies very close ($\approx 20$\,pkpc) to an \ion{O}{i} absorber at $z\approx6$ \citep[][]{Wu2021, Wu2023}.
Nevertheless, our understanding of the origin of metal absorbers and their hosts remains vague. Radiation hydrodynamical simulations are the ideal tool to unveil the nature of absorbers and their environment, guiding expectations for future observations.

Recently, fully-coupled radiation hydrodynamic simulations aimed at providing an accurate description of the EoR have become increasingly common, and large simulations like CoDa \citep[][]{Ocvirk2016}, CROC \citep[][]{Gnedin2014}, \textsc{thesan} \citep[][]{Kannan2022, Garaldi2022, Smith2022, Garaldi2024} are complemented by smaller ones such as Sphinx \citep[][]{Rosdahl2018} and SPICE \citep{Bhagwat2024}.
Based on different physics models, the current simulations reproduce the process of reionization, each with unique insights and differences, not yet reaching a concordance about how to simultaneously reproduce different observations constraining multiple aspects of reionization.
Models attempting to reproduce metal ion abundances often struggle at simultaneously capturing the evolution of both low and high-ionisation species, for which a careful calibration of feedback \citep[][]{Keating2016} and radiation is needed \citep[][]{Finlator2018, Finlator2020}.
They broadly agree that gas enrichment at $z>5$ proceeded very rapidly, with low-ionised elements tracing high-density regions close to galaxies and highly-ionised species representing an excellent probe for gas at low density \citep[][]{Oppenheimer2009, Garcia2017metals, Doughty2022}.
A correlation between galaxies and absorbers is expected, and even if different simulations produce slightly different results, the general consensus is that numerous low-mass galaxies reside in the surroundings of absorbers \citep[][]{Oppenheimer2009, Keating2014, Garcia2017, Finlator2020}.
Recently, following a particle-tracing approach, \cite{Kusmic2024} hinted at a weak positive correlation between absorber strength and galaxy stellar mass, although a direct one-to-one connection is extremely difficult to establish.

In this study, we focus on neutral oxygen (\ion{O}{i}) absorbers, frequently observed thanks to their transition at $1302$\AA, redward of Ly$\alpha$, and used as a high-redshift proxy for neutral hydrogen, given the small difference between their ionisation energies and the charge-exchange reaction to which they are subject \citep[][]{Osterbrock1989, Oh2002}.
Earlier theoretical studies showed how these absorbers are more likely to be found in the dense gas inside dark matter haloes with typical masses of $10^9\,\text{M}_\odot$ \citep[][]{Finlator2013}, with stronger systems more likely to arise in overdense galaxy environments \citep[][]{Doughty2022}.
Furthermore, \citet{Doughty2019} demonstrated that the neutral oxygen content of galaxies decreases as reionization proceeds, even if the haloes become more metal-rich, reflecting changes in the ionisation state inside the CGM.
Overall, large-volume simulations seem to reproduce the \ion{O}{i} absorber population quite well \citep[][]{Keating2014}, with a small tendency to overproduce weak absorbers and underproduce stronger ones \citep[][]{Finlator2018, Doughty2019}.

With this paper, our aim is to investigate how the distribution of neutral oxygen changes depending on galaxy properties and redshift, exploiting the unprecedented resolution of the \textsc{thesan-zoom} simulations. 
This resolution, combined with the rich physics included {and the radiation field fully-coupled to the gas via a non-equilibrium thermochemical network, is essential to reproduce the cold gas phase more naturally than in previous simulations, where a self-shielding subgrid model is needed}.
The paper is organised as follows: in Section~\ref{sec:sims} we present the simulations used in this study, followed by a qualitative overview of \ion{O}{i} absorber properties and their relation to neutral hydrogen in Section~\ref{sec:absprop}. 
We then analyse \ion{O}{i} covering fractions in Section~\ref{sec:covf}, followed by a discussion on the comparison between observational and simulated data in Section~\ref{sec:dndx}.
Finally, we present our conclusions in Section~\ref{sec:conclusions}.
 
\section{The \textsc{thesan-zoom} simulations}
\label{sec:sims}

To explore the relationship between gas and galaxies during the EoR, we used the \textsc{thesan-zoom} simulations \citep[][]{Kannan2025}, an ensemble of high-resolution zoom-in simulations.
They consist of 14 different haloes selected at $z=3$ from the \textsc{thesan-dark-1} simulation, the dark-matter (DM) only version of the large-volume radiation hydrodynamic simulation \textsc{thesan}.
The targeted haloes are chosen to represent a variety of sources in the mass range $10^8 - 10^{13}\,\text{M}_\odot$, as indicated in \citet{Kannan2025}. 
We refer to the different zoom-in regions throughout this paper according to their masses at $z=3$ in the form mX.X, such that m8.2 is the region centred on a halo of $10^{8.2} \,\text{M}_\odot$, m8.5 has a final target mass of $10^{8.5} \,\text{M}_\odot$, and so on. We refer the reader to Table 3 of  \citet{Kannan2025} for more details.
For each halo, all dark matter particles within $4$ virial radii are traced back to their initial conditions at $z=127$, and carefully resampled at higher resolution.
However, in this work, we exclusively focus on the EoR, in the redshift range $z=5-8$. 
To exploit a larger and uniform sample, we used the \textsc{thesan-zoom} simulations with standard resolution ($4$x zoom; $m_{\rm gas} \sim 9.09 \times 10^3\,\text{M}_\odot$).
We excluded the most massive galaxy, since snapshot data are not available at all redshifts of interest, due to the lack of black hole feedback implementation in the simulations, extremely important for the evolution of massive haloes at $z<6$ \citep{Kannan2025}.
Planck $2016$ cosmology is assumed, with $H_0 = 100 h$, $h = 0.6774$, $\Omega_m = 0.3089$, $\Omega_{\Lambda} = 0.6911$, $\sigma_8 = 0.8159$ \citep[][]{PlanckCosmo2016}.
\textsc{thesan-zoom} has already served as the basis for studies of the high-redshift main sequence and burstiness \citep{McClymont2025}, galaxy-scale star-formation efficiencies \citep[SFEs;][]{Shen2025}, the imprint of reionization on galaxies \citep{Zier2025}, Population III star formation \citep{Zier2025b}, the evolution of galaxy sizes \citep{McClymont2025b}, and cloud-scale SFEs \citep{Wang2025}.

The simulations are performed with the \textsc{arepo-rt} \citep{Kannan2019} code, a radiation hydrodynamic extension of \textsc{arepo} \citep{Weinberger2020, Springel2010} with the implementation of ``node-to-node'' communication presented in \citet[][]{Zier2024}.
A quasi-Lagrangian method is employed to solve hydrodynamic equations, using the characteristic Voronoi cells to represent gas elements.
Star formation follows a stochastic approach, with only self-gravitating Jeans unstable gas cells with densities above $10\,\text{cm}^{-3}$ considered eligible to form stars.
Each individual star particle represents a stellar population, typically with a mass $\approx 10^4\,\text{M}_\odot$ and an assumed Chabrier initial mass function \citep[]{Chabrier2003} with minimum and maximum stellar masses set to $0.1\,\text{M}_\odot$ and $100\,\text{M}_\odot$, respectively.
Supernovae (SNe) feedback and stellar winds are implemented using an approach based on the Stars and MUltiphase Gas in GaLaxiEs (SMUGGLE) model \citep[]{Marinacci2019}, as described in \citet{Kannan2020, Kannan2025}.
Briefly, stars with mass $M_\ast > 8\,\text{M}_\odot$ explode as SN type II, and energy, momentum and mass are injected into the closest gas cells together with metals and dust \citep{Vogelsberger2013, McKinnon2016}.  
The simulations also reproduce feedback from SN type Ia and stellar winds from O-B and asymptotic giant branch stars, thought to have a significant impact before the explosion of the first SNe.
Moreover, early stellar feedback, which has the effect of disrupting molecular clouds, is added to regulate stellar feedback before the first SNe explosions, injecting momentum for the first $5$\,Myr after star particles are formed.
{Star-forming gas, unlike the parent \textsc{thesan} simulation, is not bound to follow an effective equation of state \citep[][]{SpringelHernquist2003}. Instead, the multi-phase interstellar medium (ISM) is reproduced thanks to the high resolution of our zoom simulations and the feedback implementation applied.}

The radiation content of each cell is described across seven bins of frequency: infrared ($0.1$ - $1$ eV), optical ($1$ - $5.8$ eV), far ultra violet ($5.8$ – $11.2$ eV), Lyman–Werner band ($11.2$ – $13.6$ eV), hydrogen ionizing radiation ($13.6$ – $24.6$ eV), \ion{He}{i} ionizing radiation ($24.6$ – $54.4$ eV), and the \ion{He}{ii} ionizing radiation ($54.4$ – $\infty$ eV). 
The number of photons per cell in each frequency bin is derived from the superposition of the radiation originating outside the high-resolution region, evaluated from the parent \textsc{thesan} simulation, and the radiation generated by stars in zoom-in regions.
The latter is modelled using the BPASS stellar population synthesis model of \citet{Eldridge2017} that accounts for binary star evolution.
The evolution of the radiation field and its interaction with gas is handled using a moment-based approach \citep[]{Levermore1984}, with the introduction of a reduced speed of light, $\tilde{c} = 0.01\,c$ (where $c$ is the speed of light in vacuum), and the sub-cycling of each hydrodynamical time step $16$ times to accurately evaluate changes in the radiation field.
Radiation is coupled to gas, in particular, the simulations track different states of hydrogen (\ion{H}{i}, \ion{H}{ii}, H$_2$) and helium (\ion{He}{i}, \ion{He}{ii}, \ion{He}{iii}), used to compute primordial elements cooling.
Dust is also implemented with an updated version of the model in \citet{McKinnon2016,McKinnon2017}, with dust abundance tied to the mass of five chemical species (C,O, Mg, Si and Fe).  
Thanks to their high resolution, the \textsc{thesan-zoom} simulations are particularly well-suited for studying the gas around galaxies, in fact, the description of cooling and heating of the gas at low temperatures ($T\lesssim10^4$ K) is implemented also considering molecular hydrogen, dust-gas collisions, and metal lines, for details see \citet{Kannan2025}.
The accurate cooling description in the simulations allows for the exploration of the cold gas phase, which is fully self-consistent in the parent \textsc{thesan} simulations.
Throughout the paper, we specify whether distances are in physical (e.g., pkpc) or comoving (e.g., ckpc) units.

\section{\ion{O}{i} absorbers in simulations}
\label{sec:absprop}

\begin{figure*}
    \centering
    \includegraphics[width=\textwidth]{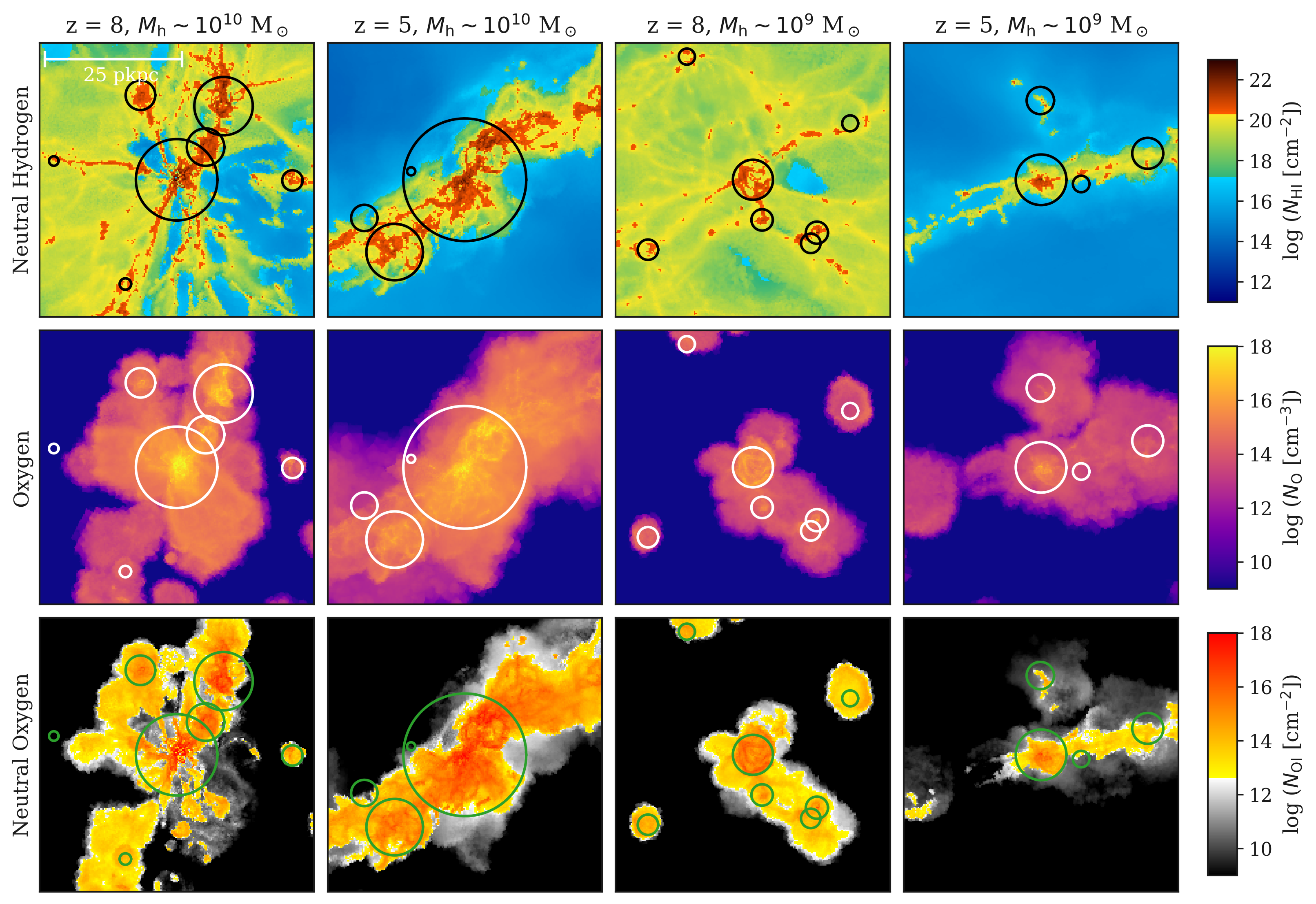}
    \caption{Column density maps of neutral hydrogen ($N_{\rm HI}$), total oxygen ($N_{\rm O}$), and neutral oxygen ({$N_{\ion{O}{i}}$}) of four different areas of size $50\times50$\,pkpc centred on haloes extracted from the m11.9 and m10.4 zoom-in regions.
    The first two columns are centred on two different haloes of mass $M_h \approx 10^{10}\,\text{M}_\odot$ at redshift $z=8$ and $z=5$, respectively, while the third and fourth columns represent regions around haloes of mass $M_h \approx 10^{9}\,\text{M}_\odot$ at the same redshifts.
    In the first row, different colours highlight different types of \ion{H}{i} absorbers, with damped Ly$\alpha$ Absorbers (DLAs) in red and Lyman Limit Systems (LLSs) in green.
    In the third row, where the neutral oxygen column density is displayed, areas with \ion{O}{i} absorbers with $N_{\rm \ion{O}{i}} > 10^{13}$ cm$^{-2}$ along the line of sight are coloured.
    In all panels, circles mark the virial radii of haloes more massive than $10^8\,\text{M}_\odot$. 
    Overall, we notice how \ion{O}{i} traces the overlap between O and \ion{H}{I}.}
    \label{fig:coolfig}
\end{figure*}

\subsection{Modelling \ion{O}{i}}

The \textsc{thesan-zoom} simulations track the abundances of seven heavy elements --- C, N, O, Mg, Ne, Si and Fe --- together with H and He. 
The non-equilibrium abundances of H and He ions are evaluated on-the-fly, as these are coupled to the radiation field in the simulations, however the simulation does not solve for the non-equilibrium chemistry of metal ions.
Combining these pieces of information together, we proceed with the estimation of the neutral oxygen density in each gas cell.

Neutral oxygen is one of the most commonly observed elements in the IGM, with its strongest line at $1302$ \AA.
It has been recognised as one of the best neutral hydrogen tracers \citep[]{Oh2002} since its first ionisation energy of $E_{\rm ion} = 13.618$ eV is only $\Delta E \approx 0.02$ eV higher than the hydrogen one.
Moreover, oxygen and hydrogen are subject to charge-exchange reactions \citep[]{Osterbrock1989} such that:
\begin{equation}
\label{eq:chargeexc}
    O^0 + H^+ \leftrightarrow O^+ + H^0 \, .
\end{equation}
In particular, given the slightly different cross-sections between the two reactions in Eq.~(\ref{eq:chargeexc}), the following relation for the neutral oxygen density holds true
\begin{equation}
    \frac{n_{\ion{O}{I}}}{n_{\ion{O}{II}}} = \frac{9}{8} \frac{n_{\ion{H}{I}}}{n_{\ion{H}{II}}} \exp{\left( -\frac{\Delta E}{kT}\right)} \, ,
\end{equation}
where $n_X$ represents the number density of the element $X$, $T$ is the temperature of the gas, $k$ is the Boltzmann constant and $\Delta E = 0.02$ eV is the difference between the ionisation energy of oxygen and hydrogen. 
At temperatures $T \gg 2.32 \times 10^2$ K, $kT \gg \Delta E$ and therefore we can assume that
\begin{equation}
\label{eq:HOeq}
    \frac{n_{\ion{O}{I}}}{n_{\rm O}} \approx \frac{n_{\ion{H}{I}}}{n_{\rm H}}
\end{equation}
considering the abundances of higher ionized oxygen states are negligible when compared to $n_{\rm{\ion{O}{i}}}$, which might not be accurate at very high temperatures.
In our study, we assume Eq.~(\ref{eq:HOeq}) holds at all temperatures, which might lead to an overestimation of neutral oxygen abundance at low temperatures ($T\approx10^2$ K).
However, temperatures below $10^4$ K are typically reached thanks to the presence of molecular gas and are mostly confined to galactic centres.
In contrast, the CGM and IGM which are the focus of our investigation, are generally hotter, with temperatures at or above $10^4$ K.
In future work, limitations arising from this assumption could be addressed by explicitly coupling the radiation field with metal abundances and other gas properties, leading to a more accurate description of the neutral oxygen content in the simulations.

\subsection{Properties of \ion{O}{i} absorbers}

We adopted Eq.~(\ref{eq:HOeq}) to determine the abundance of neutral oxygen in each gas cell within the simulation and found that, at the redshifts under investigation, \ion{O}{i} absorbers are predominantly located in the vicinity of dark matter haloes. 
To better illustrate the spatial distribution of \ion{O}{i} absorbers and their connection with oxygen enrichment and the overall ionisation state of the gas, we show in Fig.~\ref{fig:coolfig} column density maps of neutral hydrogen ($N_{\rm HI}$), total oxygen ($N_{\rm O}$) and neutral oxygen ($N_{\ion{O}{i}}$).
These maps focus on four distinct regions in the simulations, covering overdense and underdense areas at $z=8$ and $z=5$.
Specifically, the first two columns depict the gas properties within a $50$\,pkpc cube centred on two haloes of mass $M_h \approx 10^{10}\,\text{M}_\odot$ at redshift $z = 8$ and $z = 5$, respectively. 
Similarly, the third and fourth columns show the same properties for regions surrounding haloes of lower mass, $M_h \approx 10^{9}\,\text{M}_\odot$.

In all panels, the column densities are evaluated by defining a grid with cell size of $0.25$\,pkpc, over which we projected the sum of the density of the different elements.
Additionally, we checked that the column density distribution converges when considering cells smaller than $1$\,pkpc.
In the first row, we show $N_{\rm HI}$ values, with red regions indicating potential Damped Lyman-$\alpha$ Absorbers (DLAs) and green regions Lyman Limit Systems (LLSs). 
The coloured circles mark the virial radii of haloes more massive than $10^8\,\text{M}_\odot$, defined as the radius over which the average dark matter overdensity is $\rho/\bar{\rho} = 200$.
These maps reveal how at $z = 8$, regions around more massive haloes exhibit a higher number of larger ionized pockets of gas, represented in blue, which extend inside their virial radii, while high \ion{H}{i} column densities primarily trace the filamentary structure connecting haloes.
By $z=5$, however, the gas in both reproduced regions appears fairly similar, with LLSs and DLAs that can be observed exclusively in filaments, while the rest of the IGM is mostly ionized.
These findings suggest that reionization starts earlier in the gas surrounding high-mass haloes, probably because of the stronger local UV emission, but by $z=5$ it can be considered concluded in both the regions under investigation.
Overall, as pointed out by \citet{Zier2025}, the presence of local radiation field results in a more gradual and inhomogeneous reionization process compared to the outcomes predicted by a uniform UV background.

Panels in the middle row represent the total oxygen column density.
They show how even at $z=8$ metal-rich gas extends beyond the virial radii of haloes, with the highest oxygen densities concentrated near the centre of massive haloes. 
Finally, combining the data shown in the first two rows of the figure, we produced maps of the neutral oxygen column density, displayed in the third row of Fig.~\ref{fig:coolfig}. 
The coloured regions correspond to areas with $N_{\ion{O}{I}} > 10^{13}$ cm$^{-2}$, a threshold slightly below the 50\% completeness level of the latest observational surveys, typically found at equivalent widths (EW) of $0.03$ \AA \ \citep[$\log N_{\rm \ion{O}{i}}/\text{cm}^{-2} \approx 13.7$,][]{Sebastian2024}.
The distribution of neutral oxygen closely mirrors that of neutral hydrogen, as we would expect from the imposed relation in Eq.~(\ref{eq:HOeq}), except for metal-poor regions in the IGM, where even if high \ion{H}{i} column densities can be found, \ion{O}{i} absorbers are impossible to observe.
To ensure that the described trends are not specific to the selected haloes, we reproduced the same maps for the target haloes in the $13$ zoom-in regions and found no significant differences.

\begin{figure}
    \centering
    \includegraphics[width=\columnwidth]{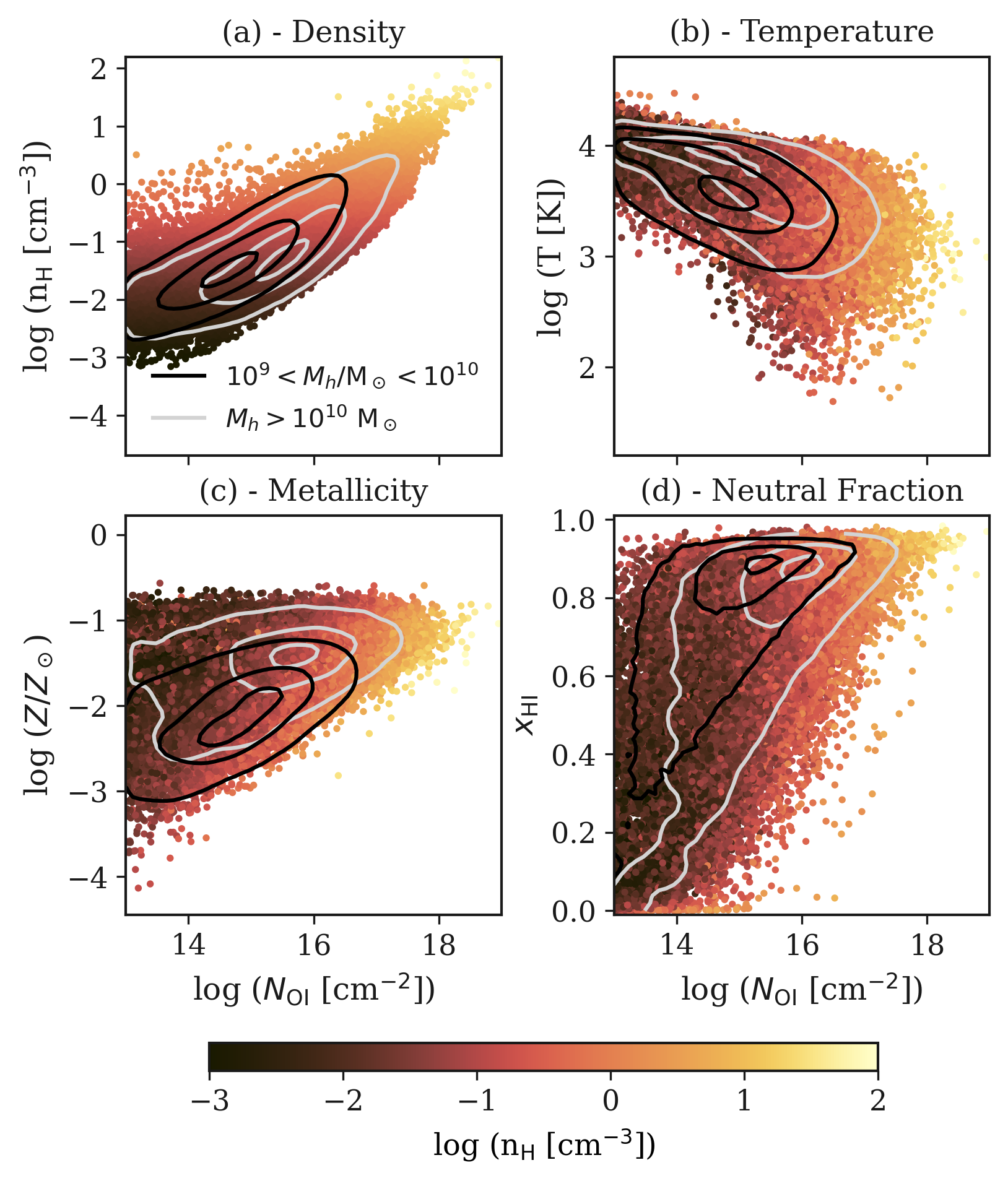}
    \caption{Gas properties of \ion{O}{i} absorbers with $N_{\ion{O}{i}} > 10^{13}$ cm$^{-2}$ inside the virial radii of haloes at $z=8$. 
    The panels show the $n_{\ion{O}{i}}$-weighted hydrogen number density ($n_\text{H})$, temperature ($T$), metallicity ($Z$), and neutral hydrogen fraction ($x_\text{HI}$) as a function of the neutral oxygen column density. 
    The points display values relative to absorbers inside the {virial radius} of high-mass haloes $M_h > 10^{10}\,\text{M}_\odot$, with the gray contours showing the density distribution of these points. The black contours mark the density distribution of absorbers inside the virial radius of lower-mass haloes ($10^9< M_h/\text{M}_\odot < 10^{10}$) which are not explicitly shown. The points are coloured according to the $n_{\ion{O}{I}}$-weighted neutral hydrogen density of absorbers. Overall, $N_{\ion{O}{i}}$ is positively correlated with $n_{\rm H}$, $Z$ and $x_{\ion{H}{i}}$, and anticorrelated with $T$.}
    \label{fig:abs_prop}
\end{figure}

\begin{figure}
    \centering
    \includegraphics[width = \columnwidth]{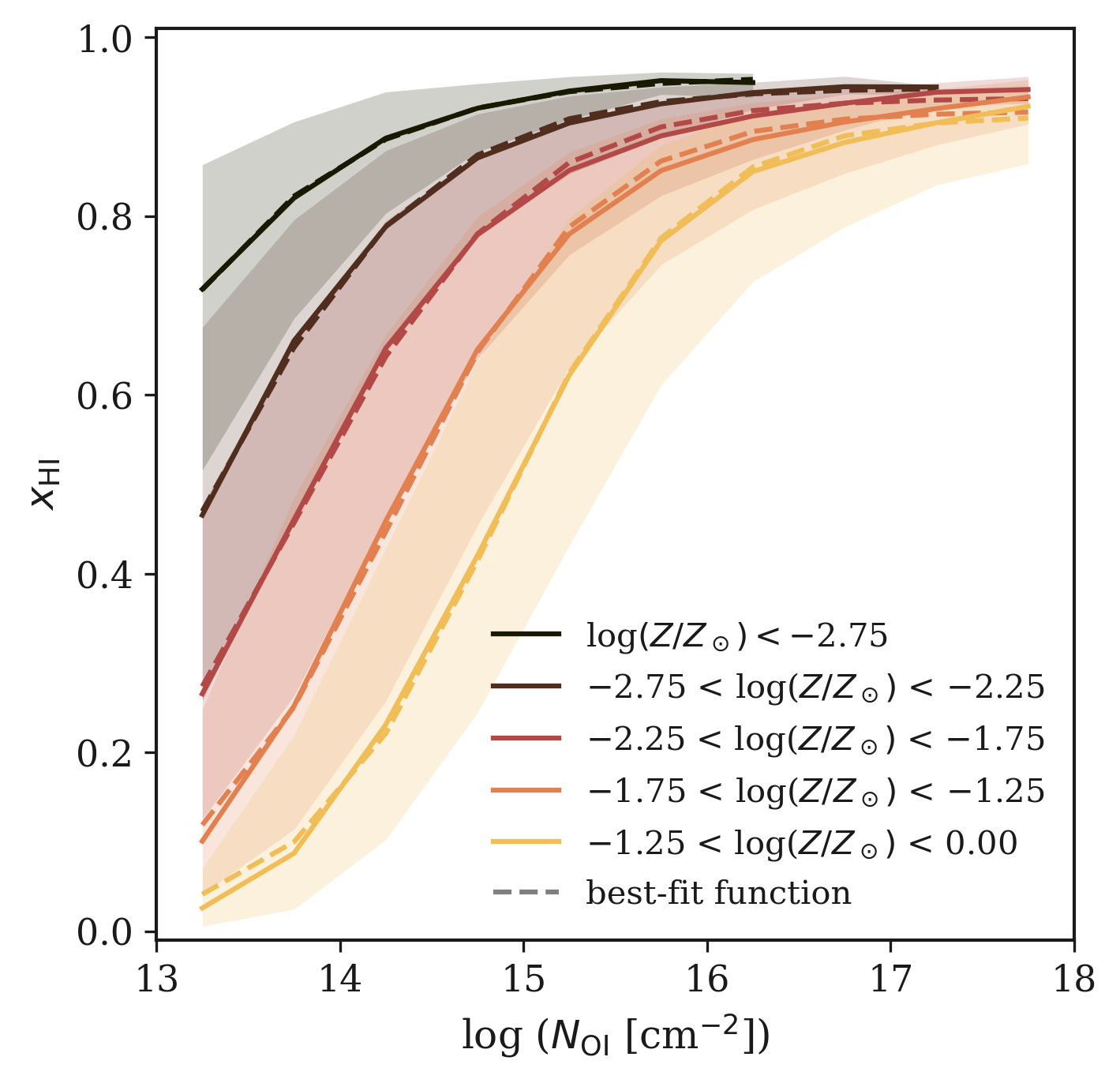}
    \caption{Relation between the neutral hydrogen fraction ($x_{\ion{H}{I}}$) and \ion{O}{i} column density of absorbers in {the virial radius of} haloes more massive than $10^9\,\text{M}_\odot$. Data are divided in $5$ metallicity bins indicated by different colours. The solid lines mark the median value of the distribution as a function of $N_{\ion{O}{I}}$, while the coloured regions identify the 16$^{\rm th}$ and 84$^{\rm th}$ percentiles. The dashed lines represent the best-fit sigmoid describing the median. The plot highlights the degeneracy between metallicity and ionisation level arising when selecting neutral oxygen absorbers with given $N_{\ion{O}{i}}$ values.}
    \label{fig:f_HI_metal}
\end{figure}

As shown in Fig.~\ref{fig:coolfig}, most of the \ion{O}{i} content of the Universe, especially absorbers with high column density values, seems to reside close to dark matter haloes, as these regions are the first to undergo chemical enrichment.
For this reason, our analysis begins with the investigation of the properties of the gas inside the virial radii of haloes.
We considered haloes residing in all the $13$ targeted zoom-in regions, such that they are entirely reproduced at high resolution.
Among these, we selected those with total halo mass $M_h > 10^9\,\text{M}_\odot$, and considering only {the virialised gas}, we evaluated the \ion{O}{i} column densities.
These are computed considering for each halo $\approx 2000$ equally spaced and parallel lines of sight crossing only gas inside the virial radius, therefore with a length that depends on the position of the line of sight with respect to the centre of the halo.
We then selected those lines of sight with $N_{\ion{O}{i}} > 10^{13}$ cm$^{-2}$ as hosts of possible absorbers.
We repeated this process across the redshift range $z = 5-8$, and investigated the gas properties of these absorbers.
Figure~\ref{fig:abs_prop} shows the relation at $z=8$ between \ion{O}{i} column density and $n_{\ion{O}{I}}$-weighted properties along the lines of sight of absorbers.
Each point identifies an absorber in massive haloes ($M_h> 10^{10}\,\text{M}_\odot$) coloured according to the value of the $n_{\ion{O}{i}}$-weighted hydrogen number density. 
Gray contour lines denote the density distribution of these absorbers, whereas the black ones show the density of absorbers in lower mass haloes ($10^9 < M_h/\text{M}_\odot < 10^{10}$), which are not explicitly shown in the figures.

In {panel (a)}, a positive correlation between hydrogen density and $N_{\ion{O}{I}}$ can be observed, with no significant variation across haloes of different masses. 
A similar positive correlation has been found in other simulations, as \citet{Oppenheimer2009} suggests.
The latter work differs from ours since they do not perform a full radiation-hydrodynamic simulation, but rather they account for fluctuations in the ionizing background in post-processing. Moreover, they apply an effective equation of state to star-forming gas in the ISM \citep[][]{SpringelHernquist2003}.
Nonetheless, the authors also find a positive correlation between the column density of low-ionised metals such as \ion{C}{ii} and the gas density.
All low-ionised absorbers in their work exhibit temperatures around $10^4$ K.
With the high-resolution cold gas phase available in \textsc{thesan-zoom}, we show in {panel (b), how temperatures for \ion{O}{i} absorbers bound to haloes vary around $10^4$ K at low column densities}, but progressively decrease with increasing $N_{\ion{O}{I}}$. 
Additionally, the scatter in temperatures broadens, spanning over two orders of magnitude at $N_{\ion{O}{I}} \approx 10^{16}$ cm$^{-2}$.
When considering lower-mass haloes, temperatures tend to be slightly lower on average, as the black contour lines suggest.
Overall, we estimate that the temperature and density of absorbers {in the virialised gas} do not change significantly with redshift, with low-$N_{\ion{O}{i}}$ absorbers typically showing lower densities and temperatures near $10^4$ K, while high-$N_{\ion{O}{I}}$ absorbers exhibit higher densities and temperatures below $10^4$ K.

{Panel (c) shows the metallicity of virialised absorbers, which spans a broad range at low column densities but converges to higher values as $N_{\ion{O}{I}}$ increases. This trend suggests the presence of an enrichment threshold proportional to the $N_{\ion{O}{I}}$ value for the formation of neutral oxygen absorbers. In our data, as the contour lines indicate, absorbers in lower-mass haloes tend to display lower metallicities on average.
As a result, achieving a given $N_{\ion{O}{i}}$ value requires a higher neutral gas fraction ($x_{\ion{H}{I}}$) compared to absorbers in higher-mass haloes.
In fact, knowing that the neutral oxygen column densities in our study can be expressed as $N_{\ion{O}{i}} = l \ n_{\ion{H}{i}} \ n_{\rm O}/n_{\rm H}$, where $l$ is the length of the line of sight, we can write the relation between neutral oxygen column density, metallicity and neutral fraction of gas as
\begin{equation}
    x_{\ion{H}{i}} = \frac{N_{\ion{O}{i}}}{l \ Z \ n_{\rm H}} \,
    \label{eq:colden_metal}
\end{equation}
where $x_{\ion{H}{i}} = n_{\ion{H}{i}}/n_{\rm H}$ and $Z \propto n_{\rm O}/n_{\rm H}$.
From this equation we can see that absorbers at lower metallicities, which are more likely to be found in lower mass haloes, as panel (c) suggests, will require a higher neutral fraction to achieve a given $N_{\ion{O}{i}}$ value.
This trend is shown in panel (d), where the black contours reach higher levels than the gray ones.
Moreover, the broad scatter displayed in $x_{\ion{H}{I}}$ values suggests that to estimate the neutral hydrogen fraction based on the \ion{O}{i} column density, additional information is required.

We therefore investigate the impact of metallicity on the relationship between neutral fraction and \ion{O}{i} column density.
We divided our absorbers in $5$ bins based on the values of $n_{\ion{O}{I}}$-weighted metallicity and plot the $x_{\ion{H}{I}} - N_{\ion{O}{I}}$ relation, as shown in Fig.~\ref{fig:f_HI_metal}.
The solid lines mark the median neutral fraction in different metallicity bins, while the shaded regions identify the $16^{\rm th}$ and $84^{\rm th}$ percentiles of the distributions. 
We find that when observing a neutral oxygen absorber, the neutral gas fraction strongly depends on the level of metal enrichment, as described in Eq.~(\ref{eq:colden_metal}).}
In fact, the same $N_{\ion{O}{I}}$ value can arise in an enriched gas with a low neutral fraction, or vice versa, in a more neutral metal-poor environment.
Moreover, we noticed that the curve defined by the median values seem to follow a sigmoid function that we therefore fit to the data. 
We show the fitting results as dashed lines in Fig.~\ref{fig:f_HI_metal} and report the best-fit parameters in the Appendix~\ref{app:best-fit_HI_NOI}.
The relation depicted in Fig.~\ref{fig:f_HI_metal} remains largely unchanged across different redshifts and it highlights one of the key challenges in using metal ion absorbers to trace neutral gas: the degeneracy between metallicity and ionisation level.

{We also note that the metallicities of our absorbers are overall lower than what presented in \citet{Finlator2013}, where the authors use an assumption similar to Eq.~(\ref{eq:HOeq}) to track neutral oxygen in the simulation. While they found $\log Z/\text{Z}_\odot \approx -0.5$ already at $z=10$, similar values are rarely present even at $z=8$ in our simulations.
However, the metallicity values displayed in their study \citep[see Fig. 2 in][]{Finlator2013} are directly derived from gas cells in the simulation, rather than being weighted by the neutral oxygen density along sightlines of absorbers, as in our case. 
Slightly higher values are therefore more likely to arise in their data.}

\subsection{The relation between \ion{O}{i} and \ion{H}{i} absorbers}

\begin{figure}
    \centering
    \includegraphics[width=\columnwidth]{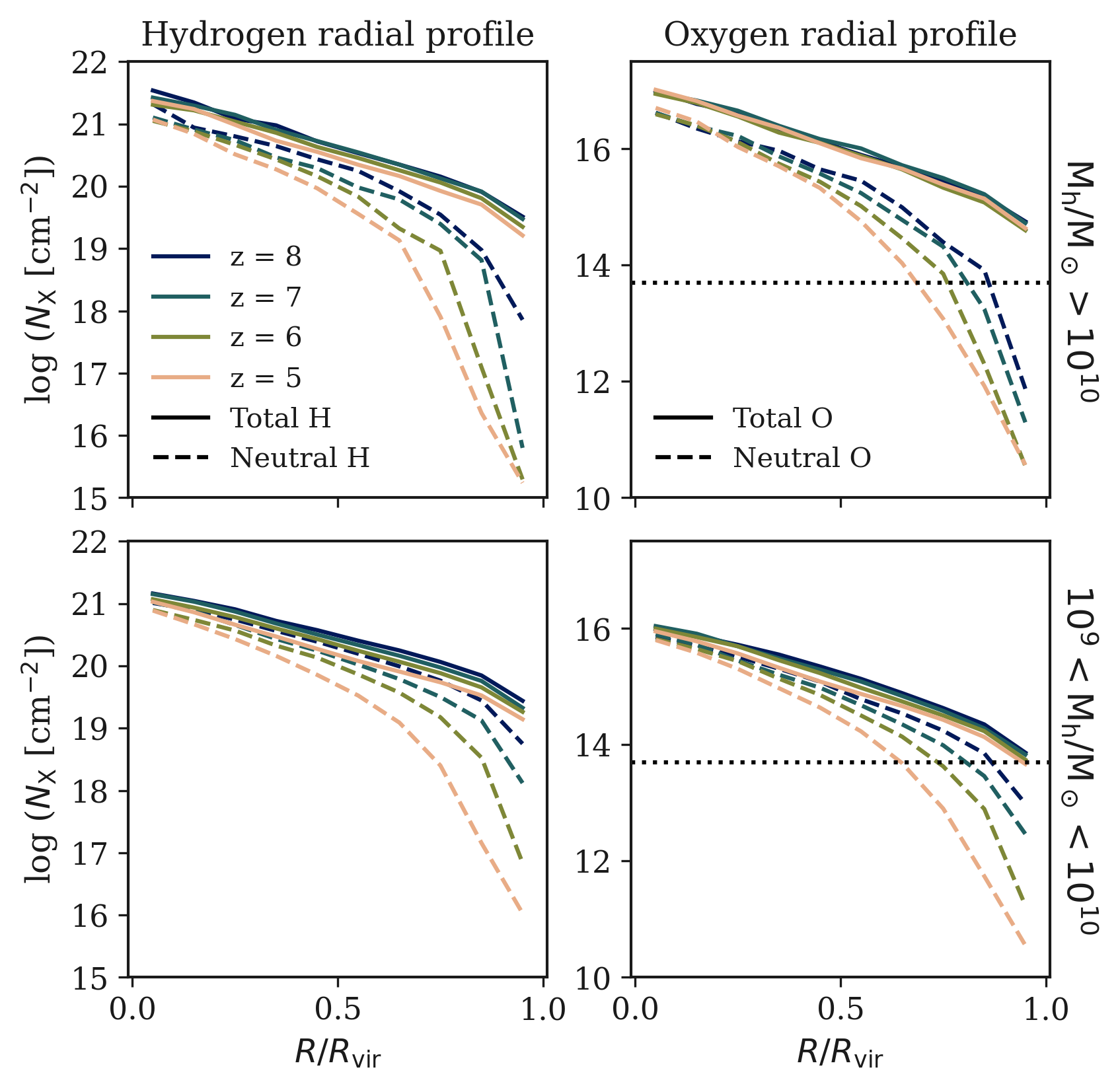}
    \caption{Median radial profile of hydrogen (left panels) and oxygen (right panels) column density values within the {virialised gas} of haloes with masses above $10^{10}\,\text{M}_\odot$ (top panels) and in the range $10^9<M_h/\text{M}_\odot<10^{10}$ (bottom panels). The solid lines trace the total hydrogen or oxygen column density and its evolution from $z=5$ to $z=8$, with darker lines marking the profiles at higher redshift and lighter ones at lower redshifts. The dashed lines delineate the profiles of neutral hydrogen or neutral oxygen column densities, following the same colour code. The dotted black line in the right panels marks the 50\% completeness level adapted from \citet{Sebastian2024}. The neutral oxygen profiles decreases as time progresses, mirroring in both high-mass and low-mass haloes the trend traced by hydrogen.}
    \label{fig:rad_prof_cd_10}
\end{figure}

Even within the {virial radius} of the same halo, \ion{O}{i} is typically not homogeneously distributed, {as can be appreciated from Fig.~\ref{fig:coolfig}}.
In Fig.~\ref{fig:rad_prof_cd_10} we present the median radial profiles of oxygen and hydrogen column densities inside the {virial radius} of high-mass ($M_h > 10^{10}\,\text{M}_\odot$, top row) and lower-mass ($10^9 < M_h/\text{M}_\odot< 10^{10}$, bottom row) haloes. 
The data are derived considering all the haloes simulated at high resolution in the $13$ zoom-in regions ($\sim 50$ high-mass haloes, $\sim 400$ lower-mass haloes at $z=7$).
For each halo, the radial profile is computed as the median value of column densities within circular annuli at fixed fractions of the virial radius.
In the plot, we represent the median of the radial profiles of similar mass haloes.

In the left panels, the solid lines indicate how the total hydrogen column density evolves with redshift, where the darkest line corresponds to the median profile at $z=8$ and the lightest at $z=5$.
The dashed lines mark the \ion{H}{i} column density profiles, following the same colour scheme. 
The plot shows how at radii larger than $0.2 \ R_{\rm vir}$, the fraction of neutral hydrogen rapidly drops with decreasing redshift, despite the total amount of hydrogen undergoing a more gradual decrease. 
This trend highlights {how the virialised gas in the outer parts of the halo} becomes progressively more ionized with decreasing redshift.
This behaviour might be caused by the combination of different effects.
First, we consider only gas inside virial radii of haloes, which grow as the Universe expands, encircling more gas at low density.
This might partly explain the decrease in neutral hydrogen, but we do not exclude that ionisation effects could also affect the change in radial profiles, as we will also discuss in Section~\ref{sec:evolofcovf} for neutral oxygen.

The plots on the right, showing total and neutral oxygen column density profiles, suggest a lack of significant evolution of the total oxygen.
Considering that the virial radius by definition is expanding as we shift to lower redshift, this behaviour outlines how gas can reach higher levels of metal enrichment at greater distances as we approach cosmic noon.
The neutral oxygen content mirrors the trend described by neutral hydrogen, {likely because of the relation imposed by Eq.~(\ref{eq:HOeq})}.
The dotted lines mark the 50\% completeness level of typical observations \citep[][]{Sebastian2024}, corresponding to $\log N_{\ion{O}{I}}/{\rm cm^{-2}} \approx 13.7$. 
On average, high column densities can be observed close to the centres of haloes, as also similar works pointed out. 
\citet{Finlator2013} shows how values of $N_{\ion{O}{I}} > 10^{16}$ cm$^{-2}$ found in regions with $r \lesssim 0.4 \ R_{\rm vir}$, are likely to arise from star-forming gas in galaxies.
In our study, however, $N_{\ion{O}{I}}$ values drop below the dotted line within the virial radius, at lower distances with decreasing redshift. 
However, {since the plots show the median profiles, it does not mean that observable absorbers are exclusively found in the centre of haloes, as also illustrated in Fig.~\ref{fig:coolfig}}. 

In \citet{Finlator2013} a similar decrease in neutral oxygen abundance in the redshift range $z=10-6$ has been pointed out, with outskirts at $r > 0.3 \ R_{\rm vir}$ being the most affected.
However, they predict a more rapid ionisation, which leaves haloes with masses similar to the ones analysed in this work with neutral oxygen fractions of $\approx 0.1 - 0.2$ at $z=6$, while our simulations suggest slightly higher average values ($\approx 0.3 - 0.4$).

\begin{figure*}
    \centering
    \includegraphics[width = \textwidth]{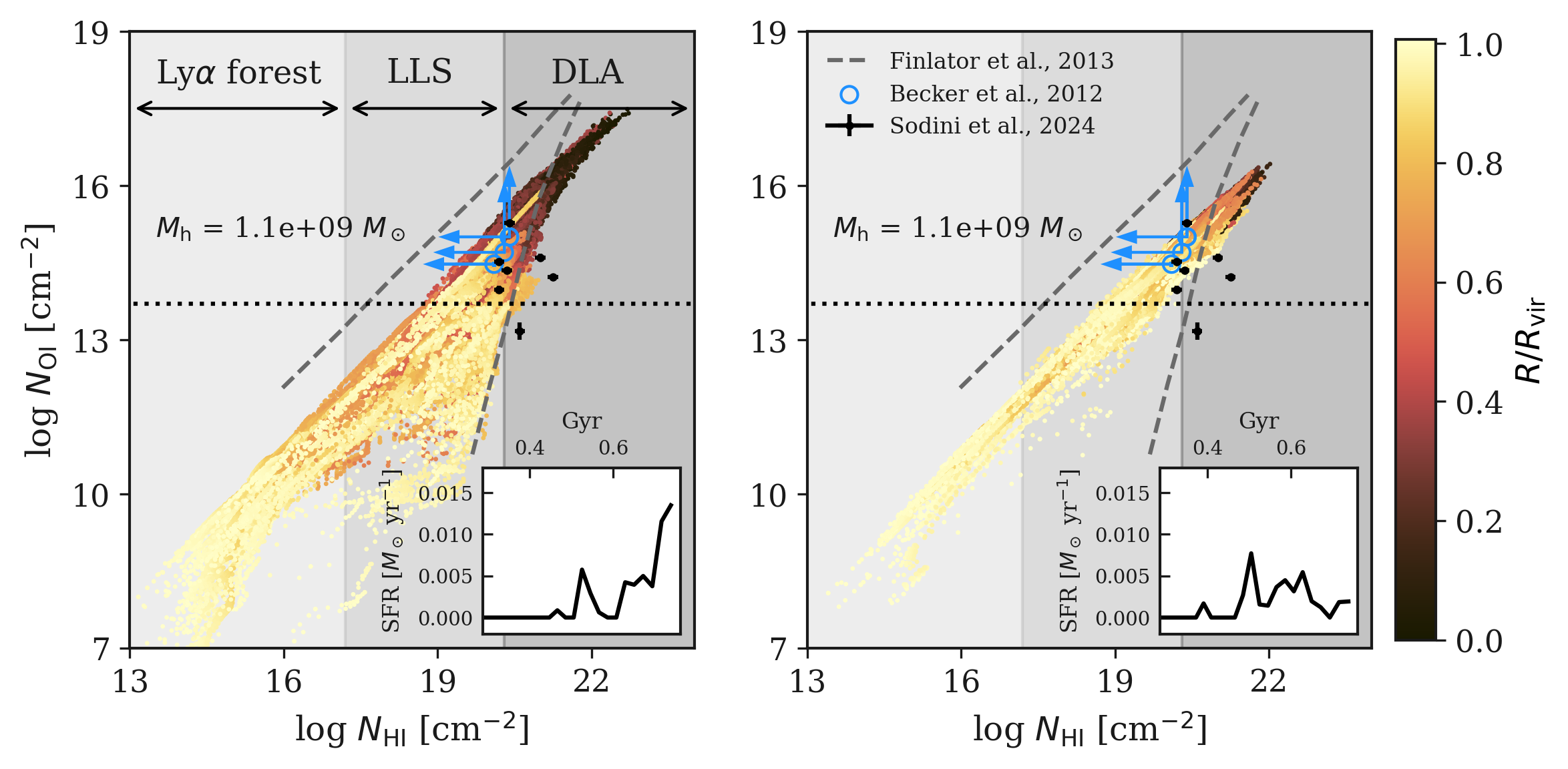}
    \caption{\ion{O}{i} and \ion{H}{i} column density relation along equally spaced and parallel lines of sight inside the {virial radius} of individual haloes selected at $z=7$. Each point represents the value along a single line of sight, coloured according to the distance from the centre of the halo. The different shades of gray in the background mark different types of neutral hydrogen absorbers. Both haloes were selected from the same simulated region, and in order to have similar masses.
    Additionally, the dashed gray lines identify the region where \citet{Finlator2013} found 99\% of their {virialised absorbers}, while the blue circles indicate observational constraints at $z\approx5$ \citep{Becker2012} and the black crosses at $z\approx6$ \citep[][]{Sodini2024}. A positive correlation between $N_{\ion{O}{i}}$ and $N_{\ion{H}{i}}$ is present, with a large scatter that might increase in some haloes. {The small inset plots show the SFH of the two haloes. Observable \ion{O}{i} absorbers are usually associated to LLS or DLAs, and higher column densities can be found in haloes with higher SFR.}}
    \label{fig:OI_to_HI}
\end{figure*}

Our analysis shows how, when using Eq.~(\ref{eq:HOeq}) to trace neutral oxygen in the simulation, \ion{O}{i} might be an excellent tracer for \ion{H}{i}, especially at high redshifts ($z>6.3$) where the Ly$\alpha$ forest is completely saturated and alternative techniques are necessary to evaluate the fraction of neutral hydrogen to constrain the reionization history.
Many observational studies have already exploited $N_{\ion{O}{I}}$ to estimate the \ion{H}{i} content of gas \citep[e.g.,][]{Cooper2019}, and in our simulations, we further investigated the relation between these two, selecting individual dark matter haloes.
Figure~\ref{fig:OI_to_HI} illustrates the relation between \ion{O}{i} and \ion{H}{i} column densities inside the {virial radius} of two distinct haloes at $z=7$, as an example, both extracted from the same simulated region and with comparable total halo mass.
The data points indicate values of column densities along $\sim 3 \times 10^4$ equally spaced parallel lines of sight going through the {virialised gas}, coloured according to their distance from the centre of the halo.

In both haloes, absorbers above $N_{\ion{O}{I}}>10^{13.7}$ cm$^{-2}$ (the level marked by the dotted line) are generally associated with LLSs or DLAs, and are more likely to be found in the innermost regions of the {virialised gas.
However, when trying to convert \ion{O}{i} to \ion{H}{i} column density, uncertainties of one order of magnitude are present, but in some cases the dispersion is even larger, as shown for example in the plot on the left.
This scatter, which cannot be fully attributed to the small difference in virial radius measuring $3.87$\,pkpc and $3.44$\,pkpc in the left and right halo respectively, suggests an inhomogeneous oxygen enrichment throughout the CGM.
We investigate the star formation history (SFH) of the two haloes up to $z=7$, since it represents one of the major contributors to metal enrichment of gas bound to the halo.
While the halo on the right began forming stars sooner, keeping a lower SFR, star formation in the left halo was delayed and proceeded with more bursty phases.
In particular, the left halo at $z=7$ is currently undergoing a starburst phase, which could explain the higher oxygen density observed in the centre of the left halo, where the high density keeps the gas neutral.
As a consequence, the upper-right region of the left plot shows a tail of black points at high column densities that are not present in the right plot.}

In general, an inhomogeneous enrichment could result from a more bursty and young SFH inside the halo, which is not uncommon to observe in the \textsc{thesan-zoom} simulations, especially in low-mass haloes at high redshift \citep[][]{McClymont2025} with implications for the excess of UV-bright galaxies observed by JWST~\citep[e.g.,][]{Shen2023}.
However, we do not exclude that also the galactic environment in which the galaxy resides could play a role in enriching the CGM of nearby smaller haloes. 
In fact, we estimate that the average SFR over the last $100$\,Myr in galaxies within $50$\,pkpc from the halo in the right panel is about a factor $0.5$ higher than the one on the left. 
It may be that a higher SFR in galaxies surrounding the halo might contribute to a more homogeneous enrichment.

The large scatter in the relation between neutral oxygen and hydrogen column densities was also shown in previous studies.
In \citet{Finlator2013}, for example, the authors analysed this relation inside the virial radii of haloes more massive than $\sim 10^{7.6}\,\text{M}_\odot$.
The dashed gray lines in Fig.~\ref{fig:OI_to_HI} mark the region over which the 99\% of their data at $z=7$ lies. 
Overall, they found a similar trend, with DLAs showing higher and well-constrained \ion{O}{i} column density, while wide scatter can be found especially at $N_{\ion{O}{i}} < 10^{14}$ cm$^{-2}$.
However, their values seem to be shifted to the left, meaning that their {gas} is on average more oxygen rich, which could be due to the fact that they also consider haloes more massive than $10^9\,\text{M}_\odot$, together with a stronger level of enrichment predicted by their model, as also suggested by the comparison of the overall metallicities \citep[Fig.~2 from][]{Finlator2013}.
The blue circles show constraints from \citet{Becker2012}, that at $z\approx5$ measured values of $N_{\rm HI}$ and $N_{\ion{O}{i}}$ along quasars lines of sight.  
Furthermore, the black crosses mark the column densities found in the work of \citet{Sodini2024}, where the authors measured values of $N_{\rm HI}$ of absorbers at $z\approx6$ by analysing Ly$\alpha$ damping wings of proximate DLAs. 
These data points suggest lower enrichment values, but can still be reproduced in the {virialised gas} of haloes by our simulations, especially in environments similar to the one associated with the left halo in Fig.~\ref{fig:OI_to_HI}, but we do not exclude that these observations might probe further regions in the IGM where we expect lower metal enrichment.
However, the errors on $N_{\rm HI}$ values in \citet{Sodini2024} do not account for uncertainties in modelling the Ly$\alpha$ region of the continuum, and are likely to be larger.
In conclusion, even if we agree with previous observations and simulations \citep[e.g.,][]{Finlator2013, Becker2012} on the fact that absorbers with column densities above the 50\% level of completeness are associated with LLS or DLA systems, we argue that multiple factors including star formation inside the halo and in nearby galaxies could introduce a large scatter in the $N_{\ion{O}{I}}-N_{\ion{H}{I}}$ relation.

\section{\ion{O}{i} covering fraction around haloes}
\label{sec:covf}

\begin{figure*}
    \includegraphics[width = \textwidth]{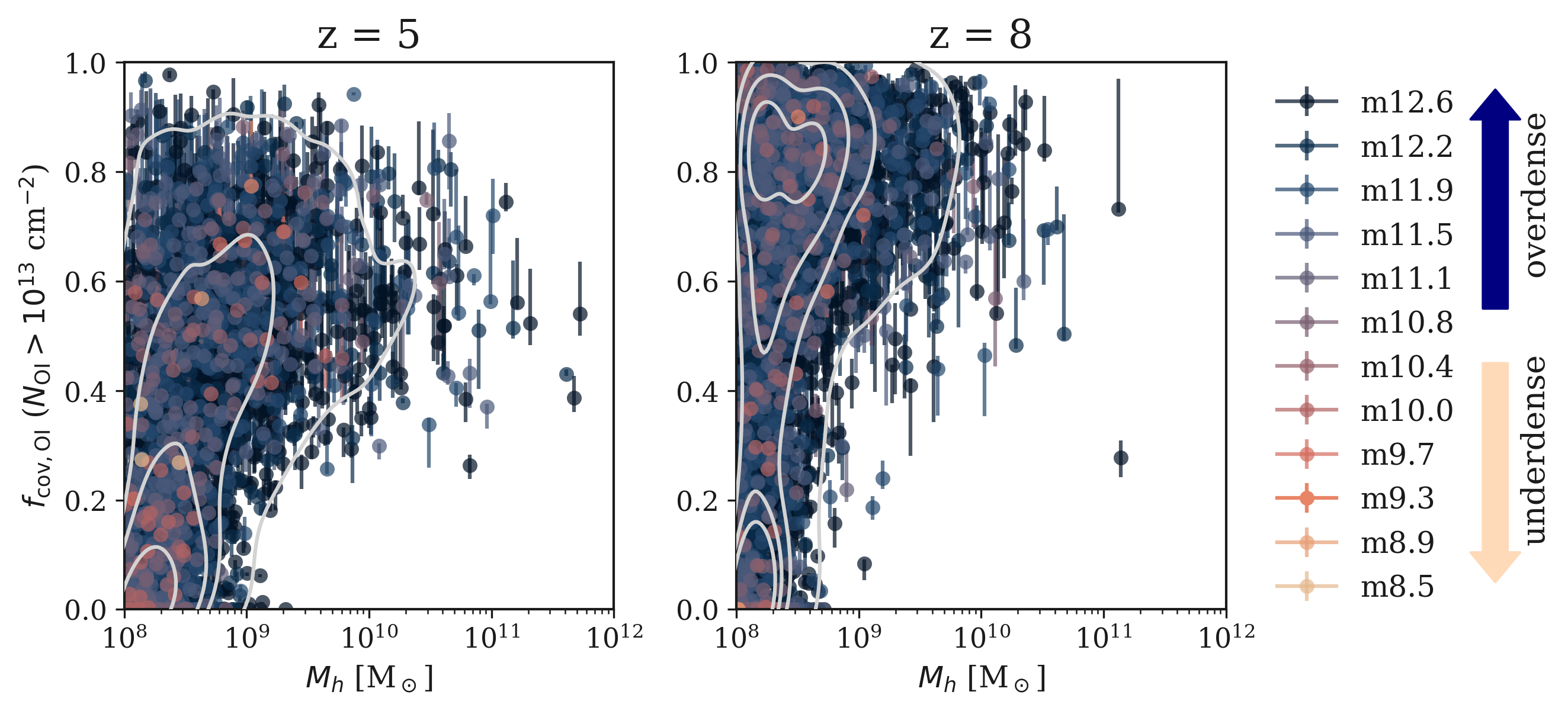}
    \caption{\ion{O}{i} covering fraction ($f_{\rm cov, OI}$) as a function of halo mass at $z=5$ and $z=8$. For each halo three covering fractions have been evaluated observing the halo from three perpendicular directions. The median covering fraction of each halo is represented by a filled circle, while error bars mark the minimum and maximum values for the same halo. Each point is coloured according to the zoom-in region to which it belongs, with darker colours representing more overdense areas. Gray contour lines are used to highlight the density of data points. Despite the large scatter displayed by the data, an overall decrease of covering fractions with decreasing redshift can be observed.}
    \label{fig:covf_halomass}
\end{figure*}

\begin{figure}
    \centering
    \includegraphics[width=\linewidth]{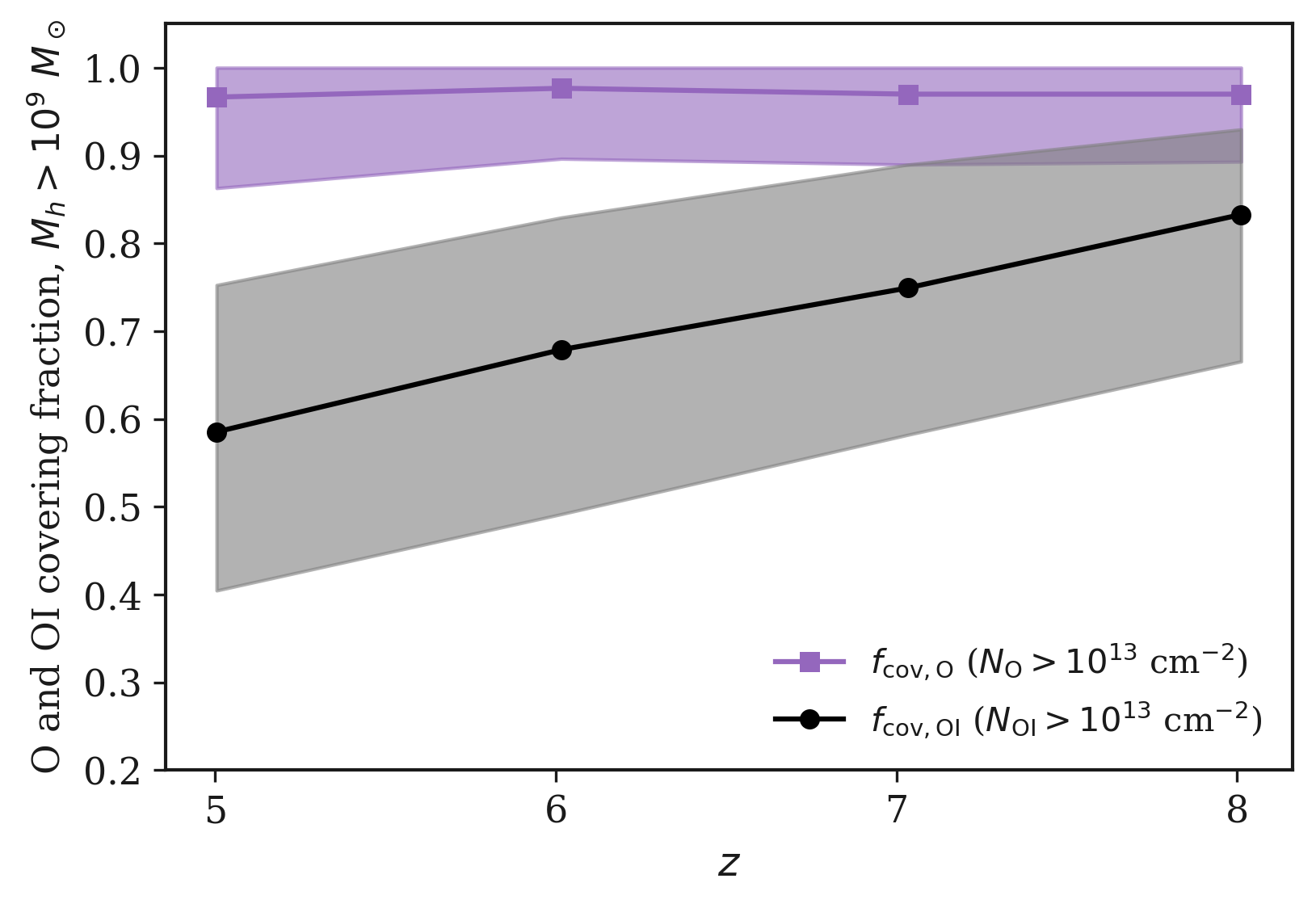}
        \caption{Distribution of covering fractions of \ion{O}{i} (black dots) and total \ion{O}{} (purple squares) for haloes with $M_h \geq 10^9\,\text{M}_\odot$, evolving from $z=5$ to $z=8$.
        Both covering fractions are computed selecting a column density threshold of $10^{13}$ cm$^{-2}$. The median of the distribution is marked by solid lines, while the shaded regions indicate the area between the $16^{\rm th}$ and $84^{\rm th}$ percentiles. The ionisation state of the gas inside the virial radius changes, becoming more ionised at lower redshifts.}
    \label{fig:covf_median_z_ev}
\end{figure}

Most observational studies \citep[e.g.,][]{D'Odorico2013, D'Odorico2023, Sebastian2024} explored absorber properties through quasar absorption spectroscopy.
This process is usually reproduced in theoretical studies by shooting random lines of sight throughout the simulated volume \citep[e.g.,][]{Finlator2015}.
In zoom-in simulations, because of the limited high-resolution area, the same procedure would inevitably lead to biased results. 
Therefore, we address the study of the neutral oxygen distribution and its connection to galaxies starting from the analysis of the gas around each halo, introducing the \ion{O}{i} covering fraction ($f_{\rm cov, OI}$), a tool also used in similar previous studies \citep[][]{Finlator2013, Doughty2019}.
This is defined as the portion of the projected area inside the virial radius of a halo covered by lines of sight along which the \ion{O}{i} column density is higher than a fixed threshold. 

For each high-resolution halo with total mass above $10^8\,\text{M}_\odot$, we measured values of $N_{\ion{O}{I}}$ along $\sim 3 \times 10^4$ parallel and equally spaced lines of sight {crossing gas in the virial radius}.
This process is repeated along three fixed perpendicular directions, obtaining three values of neutral oxygen covering fractions for each halo.
Figure~\ref{fig:covf_halomass} illustrates the relationship between covering fraction and halo mass, when selecting a column density threshold of $N_{\ion{O}{I}} = 10^{13}$ cm$^{-2}$, such that for each halo the median covering fraction is represented by a circle, while the associated error bars identify the minimum and maximum values. 
The colours denote the zoom-in region to which the haloes belong, with darker points marking haloes residing in denser large-scale environments, i.e., zoom-in regions that form higher-mass haloes at $z=3$.
No points associated with the low-density region m8.2 are present, since no haloes above the selected mass threshold at $z\geq5$ are found.
However, the lack of any significant difference between points of different colours suggests that the large-scale environment plays a minor role in determining the value of the covering fraction.

The scatter displayed by the data points in Fig.~\ref{fig:covf_halomass} is extremely large.
The dispersion is especially wide in very low-mass haloes ($M_h \lesssim 10^9\,\text{M}_\odot$), while higher-mass haloes ($M_h \gtrsim 10^9\,\text{M}_\odot$) all exhibit values above $0.1$.
Despite the presence of this large dispersion, an overall decreasing trend in covering fraction values with redshift is highlighted by the gray contour lines in the plots.
In fact, $f_{\rm cov, OI}$ values close to unity, which are present at $z=8$, become extremely rare at $z=5$.

\subsection{Redshift evolution of \ion{O}{i} covering fractions}
\label{sec:evolofcovf}

Motivated by the observation above, we present in Fig.~\ref{fig:covf_median_z_ev} the redshift evolution of the \ion{O}{i} covering fraction for high-mass haloes.
Specifically, since we evaluated the covering fractions along three perpendicular directions, we associate to each halo a covering fraction probability density defined by a Gaussian function with a mean and standard deviation derived from the three \ion{O}{i} covering fractions values. 
By summing the probability density functions associated to haloes with mass $M_h > 10^9\,\text{M}_\odot$, we obtain at each redshift a \ion{O}{i} covering fraction distribution.
The evolution of the median and $16^{\rm th}$ and $84^{\rm th}$ percentiles of this distribution are shown in Fig.~\ref{fig:covf_median_z_ev}, when selecting a column density threshold of $10^{13}$ cm$^{-2}$ (black circles and shaded region).
Larger \ion{O}{i} covering fractions are more likely to be observed at high redshift ($z=8$) than at the end of reionization ($z=5$).
This suggests that, within the virial radius, the neutral oxygen column density is on average higher at higher redshift, consistent with the trend observed in the evolution of the radial profiles in Fig.~\ref{fig:rad_prof_cd_10}.
A similar behaviour is shown when a higher column density threshold of $10^{14}$ cm$^{-2}$ is adopted, with average values of covering fractions approximately $0.15$ lower.
A declining neutral oxygen density with decreasing redshift is also found in observations of metal ion absorbers \citep[][]{Becker2019}, which suggest a similar evolution in the ionisation state.
However, the incidence of absorbers is not directly comparable to the covering fraction, as it also depends on the halo mass function and it accounts for the presence of neutral oxygen in the IGM outside the virial radius. 
{For haloes with $M_h < 10^9\,\rm{M}_\odot$ we point out that despite the large scatter}, an overall decrease in the \ion{O}{i} covering fraction can be appreciated from Fig.~\ref{fig:covf_halomass}. This is probably also due to reionization, that in these low-mass haloes has been shown to cause a drop in the baryonic mass between $z=7$ and $z=5$ \citep[][]{Zier2025}.  

A similar redshift evolution of \ion{O}{i} covering fractions has been reported in the previous work of \cite{Finlator2013}, where the decrease was observed to be stronger, dropping from average values of $\approx 0.7$ at $z=10$ to $\approx 0.2$ at $z=6$. Additionally, their data points exhibit a smaller scatter. 
\citet{Doughty2019} also observed a comparable trend, noting that the decline in covering fractions for higher-mass haloes mainly occurs between $z=8$ and $z=7$, while in lower-mass haloes it is delayed to redshift $7-6$.
Our results for high-mass haloes ($M_h > 10^{10} \,\text{M}_\odot$) align with these findings, since nearly half ($\sim 49\%$) of the total decrease in covering fraction occurs between $z=8-7$.
However, for lower-mass haloes ($10^9 < M_h/{\rm M_\odot}<10^{10}$), our simulation predicts a later transformation, with the most significant decrement in \ion{O}{i} covering fraction ($\sim 41\%$ of the total) happening between $z=6 - 5$.
This difference may be partially caused by a slightly earlier reionization in \textsc{Technicolor Dawn}, the simulation used in \citet{Doughty2019}, where neutral hydrogen fractions of $\sim 1\%$ are reached at $z=6$ \citep{Finlator2018}.
For comparison, in the \textsc{thesan-1} simulation, a neutral hydrogen fraction of $\sim 7\%$ is reached at the same redshift \citep[][]{Kannan2022}. 
Moreover, in the two simulations, differences in the galaxy formation models and gas resolution might cause different neutral oxygen distributions.
A more detailed comparison with the values shown in \citet{Doughty2019} is challenging, since in their work covering fractions are computed selecting a fixed radius of $500 \ h^{-1}$ ckpc, which translates to $\approx 120$\,pkpc at $z=5$, a factor $>10$ larger than the virial radii of our haloes. 
As a result, they generally obtain lower values of covering fractions, with none of their data points above $0.4$.

To explain the decrease in the \ion{O}{i} covering fraction with redshift, it is essential to examine both the metal enrichment and the ionisation state of the gas.
As already shown in Fig.~\ref{fig:rad_prof_cd_10}, the radial profile of the total oxygen column density remains constant across the examined redshift range. 
For this reason, we believe that the change in \ion{O}{i} covering fraction is primarily driven by a change in ionisation state.
To test our hypothesis, we repeated the same analysis, focusing on the total oxygen in the virialised gas.
We evaluated the total O covering fractions of the same haloes as the portion of area inside the virial radius covered by lines of sight displaying total oxygen column densities higher than $N_{\rm O} = 10^{13}$ cm$^{-2}$, and repeated the same procedure.
We show the evolution of the oxygen covering fraction distribution with the purple squares and shaded region in Fig.~\ref{fig:covf_median_z_ev}.
The total O covering fraction does not show a significant evolution in the redshift range analysed.
This trend, valid also when selecting the higher column density threshold of $10^{14}$ cm$^{-2}$ but with values that are about $0.15$ lower, suggests that on average in haloes more massive than $10^9\,\text{M}_\odot$ the effects of metal enrichment compensate for the enlargement of the virial radius.
Furthermore, we tested this result by considering spheres with fixed radius of $5$\,pkpc around the same haloes, and as expected, we found that the total oxygen covering fraction increases as the redshift decreases.
Therefore, we conclude that the decrement in the neutral oxygen covering fraction with decreasing redshift is mainly due to a change in the ionisation state of the gas, {as already suggested by Fig.~\ref{fig:rad_prof_cd_10}}.
We identified the ionising radiation flux generated both by stars residing in the halo and by the overall galactic environment as the main cause of this transformation. 
However, a more detailed analysis is required to quantify the relative contribution of each component, which will be the focus of future work.
Although we do not exclude heating from SN feedback to also play a role, we found that only a small percentage of haloes present a significant fraction of ionized gas at temperatures above $10^6$ K.

\subsection{Explaining the \ion{O}{i} covering fraction scatter}

\begin{figure}
    \centering
    \includegraphics[width=\columnwidth]{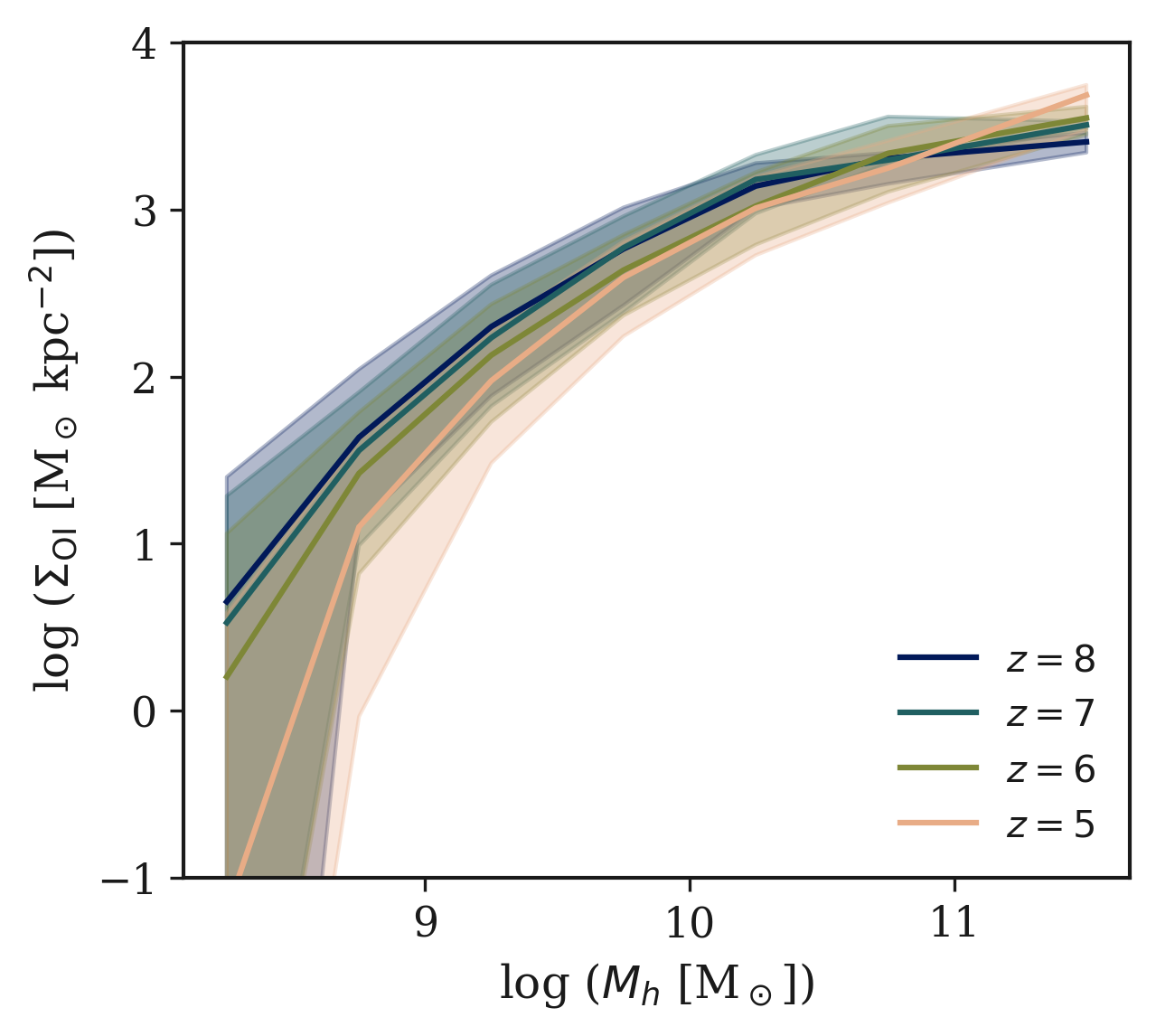}
    \caption{Evolution of the median neutral oxygen surface density within the virial radius ($\Sigma_{\ion{O}{I}}$) as a function of the halo mass, between redshift $z=5$ and $z=8$. The median of the distribution is marked by solid lines, while shaded regions identify the area between the $16^{\rm th}$ and $84^{\rm th}$ percentiles. Darker (lighter) colours are associated to higher (lower) redshifts. The neutral oxygen surface density shows a clear positive correlation with halo mass without significant evolution with redshift, especially for M$_h > 10^9\,\text{M}_\odot$.}
    \label{fig:sigma_mass}
\end{figure}

\begin{figure*}
    \includegraphics[width=\textwidth]{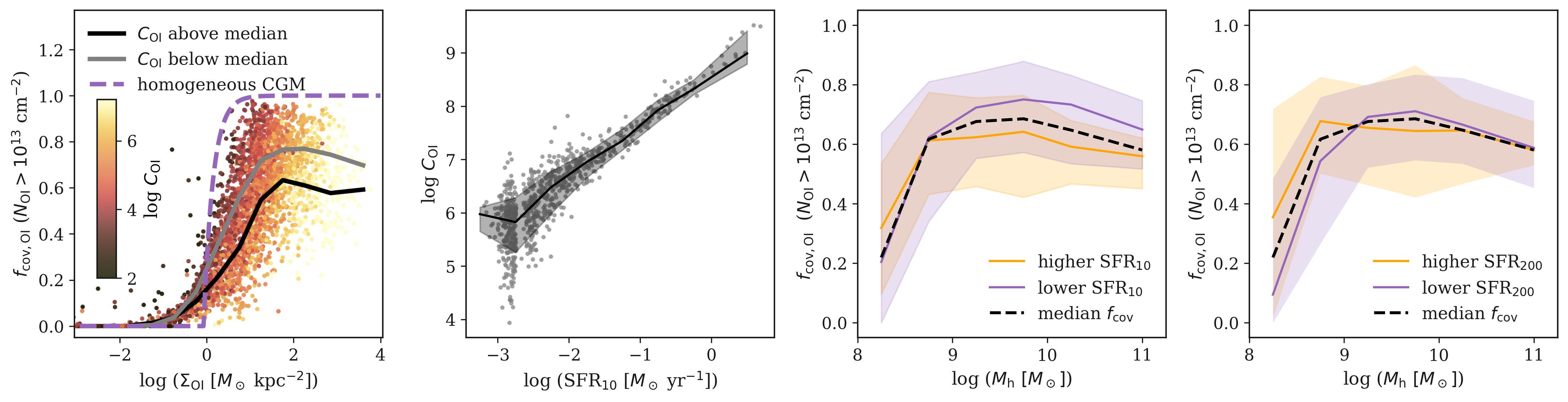}
    \caption{The role of clumpiness and SFR in determining the \ion{O}{i} covering fraction at $z=6$.
    From left to right: (\textit{i}) \ion{O}{i} covering fraction as a function of the \ion{O}{i} surface density of haloes.
    The dashed purple line marks the expected relation if \ion{O}{i} was homogeneously distributed within the {virial radius} of haloes, as demonstrated in Appendix~\ref{app:covfrac_theo}.
    Each point represents the median $f_{\rm cov, OI}$ of haloes when selecting a column density threshold of $10^{13}$ cm$^{-2}$.
    Colours indicate the clumping factor value ($C_{\ion{O}{i}}$), such that darker points have a more homogeneous distribution than lighter points. 
    The black and gray lines trace the median \ion{O}{i} covering fraction when considering haloes with the highest (above the median), and lowest (below the median) clumping factors as a function of surface density. 
    (\textit{ii}) Relation between clumping factor and the SFR of haloes at $z=6$ evaluated over the last $10$\,Myr. Each gray point represents a halo, while the black line traces the median $C_{\ion{O}{i}}$ at different SFR values, with the shaded region marking the $16^{\rm th}$ and $84^{\rm th}$ percentiles.
    (\textit{iii}) $f_{\rm cov, OI}$ distributions of two subsets of haloes as a function of halo mass at $z=6$. In purple those with SFR averaged over $10$\,Myr below the median evaluated in haloes of comparable mass, in orange the ones with SFR above the median. 
    Solid lines trace the median of the distributions, while the shaded regions indicate the $16^{\rm th}$ and $84^{\rm th}$ percentiles.
    The dashed black line shows the overall median of neutral oxygen covering fraction values across mass bins.
    (\textit{iv}) Same as in the third panel, but with SFR values averaged over longer timescales of $200$\,Myr, highlighting the different behaviour of low and high-mass haloes.
    Overall, at high halo masses, higher values of SFR over $10$\,Myr are associated to lower covering fractions.}
    \label{fig:covf_sigma}
\end{figure*}

Previous works using different simulations \citep[][]{Finlator2013, Doughty2019} pointed out the presence of a positive correlation between halo mass and neutral oxygen covering fraction.
However, the substantial scatter in our data, as shown in Fig.~\ref{fig:covf_halomass}, prevents the robust identification of a possible correlation. 
There is only a tentative indication at $z=5$, where the gray contour lines appear to trend toward higher covering fractions for higher-mass haloes. 
We also investigate the behaviour of neutral oxygen covering fractions when compared to other galaxy properties, finding a similar correlation and dispersion with the dark matter mass, total oxygen and neutral oxygen mass in the virial radius. 
On the other hand, a clear correlation between SFR evaluated on different timescales (from $2$\,Myr to $200$\,Myr) is more challenging to observe, probably hidden by the large dispersion present even in this case.

To explain the positive correlation observed with the halo mass and understand the reasons behind the wide scatter, we approached the concept of covering fraction from a theoretical point of view, solely based on its definition.
We consider the case in which neutral oxygen is homogeneously distributed within a spherical volume defined by the virial radius. 
Under this assumption, we find that the covering fraction is determined exclusively by the \ion{O}{i} surface density ($\Sigma_{\ion{O}{i}} = M_{\ion{O}{i}}/(\pi R_{\rm vir}^2)$). 
In fact, haloes with higher surface densities should display higher covering fractions, as we demonstrate in detail in  Appendix~\ref{app:covfrac_theo}.

By measuring the \ion{O}{i} surface density for the simulated haloes, we note that it increases with halo mass, as represented in Fig.~\ref{fig:sigma_mass}.
In this figure, solid lines trace the evolution of the median value of neutral oxygen surface density as a function of halo mass, from $z=8$ (darkest line) to $z=5$ (lightest line).
A tight positive correlation becomes particularly pronounced at halo masses above $10^9\,\text{M}_\odot$, probably reflecting the longer and more intense star formation activity hosted in high-mass galaxies, which could lead to an increase in the overall oxygen content, likely to remain neutral in high-density regions.
Since we expect from theory that the \ion{O}{i} covering fraction increases with neutral oxygen surface density in haloes with a homogeneous \ion{O}{i} distribution, and we showed how the \ion{O}{i} surface density, in turn, increases with halo mass, a positive correlation between $f_{\rm cov, OI}$ and halo mass is expected.
However, the substantial scatter displayed in Fig.~\ref{fig:covf_halomass} cannot be fully attributed to the modest dispersion in the $M_{h} - \Sigma_{\ion{O}{i}}$ relation, probably caused by slightly different star formation histories.
For this reason, we investigate the presence of possible scatter in the $\Sigma_{\ion{O}{i}} -$ \ion{O}{i} covering fraction relation, which we think might be the main cause behind the dispersion shown in Fig.~\ref{fig:covf_halomass}.

As highlighted in Appendix~\ref{app:covfrac_theo}, deviations from the predicted relation between covering fraction and surface density should be favoured by the inhomogeneous distribution of \ion{O}{i}.
We investigated the role of \ion{O}{i} clumpiness by introducing the clumping factor, defined as
\begin{equation}
    C_{\ion{O}{i}} = \frac{\left< n_{\ion{O}{i}}^2 \right>_{V}}{\left< n_{\ion{O}{i}}\right>_V^2} 
\end{equation}
where $n_{\ion{O}{i}}$ is the number density of neutral oxygen in each gas cell with centre of mass falling inside the virial radius of a halo, and $\left<\,\right>_V$ indicates the volume-weighted average.
The role of $C_{\ion{O}{i}}$ is illustrated in the first panel of Fig.~\ref{fig:covf_sigma}, where each data point in the plot indicates the median \ion{O}{i} covering fraction of haloes at $z=6$ as a function of $\Sigma_{\ion{O}{i}}$, coloured according to the value of the \ion{O}{i} clumping factor, such that darker points represent haloes in which neutral oxygen is more homogeneously distributed. 
The dashed purple curve traces the theoretical relation in haloes where \ion{O}{i} is homogeneously distributed (see Appendix~\ref{app:covfrac_theo}).
Overall, the data points show how the two variables rise together.
At low $\Sigma_{\ion{O}{i}}$, the covering fraction is predicted to be zero, as haloes in this regime lack sufficient neutral oxygen for lines of sight to exceed the column density threshold of $10^{13}$ cm$^{-2}$.
However, in this case, clumpiness can locally enhance the neutral oxygen density values, allowing some lines of sight to overcome the threshold, increasing the covering fraction. 
At higher surface densities, {where most of the haloes would show covering fractions of unity if the \ion{O}{i} distribution was homogeneous, a large scatter is displayed.}
By looking at the data points, it is hard to identify the role of clumpiness in this dispersion, as an overall trend of increasing \ion{O}{i} clumping factor with increasing surface density is present.
We therefore divided our data in $\Sigma_{\ion{O}{i}}$ bins, evaluated the median clumping factor for each bin, and identified the haloes with clumping factors above and below the median value.
The solid black and gray lines in Fig.~\ref{fig:covf_sigma} show the covering fraction of haloes in these two subsets, respectively.
The analysis reveals that, at high $\Sigma_{\ion{O}{i}}$ values, haloes with higher clumping factors display lower \ion{O}{i} covering fractions on average.

Additionally, we identified a group of outliers in the plot shown in the first panel of Fig.~\ref{fig:covf_sigma}.
There are a few scattered haloes with low surface density and low clumping factor displaying values of \ion{O}{i} covering fractions higher than expected from our theoretical relation.
These points represent small haloes in which only less than $\sim 50$ gas cells with centre of mass inside the virial radius can be found.
Because the virial radius of these haloes is very small, they are preferentially seen at higher redshift, and it is likely that the selected gas cells partially extend over the virial radius.
In this case, the average number densities displayed by the particles will be slightly lower than those of a gas cell completely contained inside the CGM, and the average surface density could be underestimated, causing a shift to the left in the plot.

Our results show that the \ion{O}{i} clumping factor is a key element to increase the scatter in the relation between covering fraction and surface density.
Therefore, processes that favour an inhomogeneous distribution of neutral oxygen in haloes might be essential to explain the wide scatter in Fig.~\ref{fig:covf_halomass}, especially at high masses ($M_h > 10^9\,\text{M}_\odot$), where the dispersion in the $\Sigma_{\ion{O}{i}} -M_h$ relation is not highly significant.
By analysing the definition of clumping factor, which depends only on the distribution and concentration of neutral oxygen, we expected it to be correlated with the star formation activity inside the halo.
In fact, a higher SFR evaluated on timescales {just long} enough in order to have significant oxygen enrichment, would fill the gas close to the star-forming region with high-density oxygen, while also emitting energetic UV photons able to ionize oxygen in the outskirts, where the density is lower.
For this reason, we analysed the relation between \ion{O}{i} clumping factor and SFR, finding a very clear positive correlation, especially for haloes with masses above $10^9\,\text{M}_\odot$, when considering SFR averaged over short periods of time, from $2$ to $50$\,Myr, preceding the simulation snapshot.
As an example, the relation between $C_{\ion{O}{i}} - {\rm SFR_{10}}$ is represented in the second panel of Fig.~\ref{fig:covf_sigma}, where the solid line traces the median clumping factor as a function of the SFR evaluated over the last $10$\,Myr at $z=6$.
The shaded regions mark the area between the $16^{\rm th}$ and $84^{\rm th}$ percentiles, highlighting the little dispersion present. 
These values of SFR averaged on short periods also show a strong positive correlation with the average ionising photon rate inside the virial radius of high-mass haloes. 
In fact, by assuming the reduced speed of light $\tilde{c}$, in about $10$\,Myr photons would travel a distance of $\sim 30$\,pkpc, which is of the order of the virial radius of the most massive haloes ($M_h \approx 10^{11}\,\rm{M}_\odot$) at $z=5$.

To test the influence of SFR, we show in the third and fourth panels of Fig.~\ref{fig:covf_sigma} the \ion{O}{i} covering fraction distribution as a function of the halo mass when selecting two subsets of haloes.
The orange lines mark the result for haloes with SFR above the median value evaluated in a similar mass range, while the purple lines identify the haloes with SFR below the median.
In the third panel, values of SFR are obtained by averaging over $10$\,Myr before the snapshot, while in the fourth panel a longer timescale of $200$\,Myr is used. 
Focusing on masses above $10^9\,\text{M}_\odot$, the figure shows that haloes with higher SFR$_{10}$ tend to display lower values of \ion{O}{i} covering fraction, while the SFR computed on longer timescales does not seem to play a significant role.
This suggests that in these galaxies the ionising photons emitted by recent star formation activity are able to ionize oxygen, especially at the low densities that can be found in the outskirts of the halo, as seen in Fig.~\ref{fig:rad_prof_cd_10}.
This photoionization effect leads to an increase in the \ion{O}{i} clumpiness, and a following decrease in the \ion{O}{i} covering fraction.

For low-mass haloes ($10^8 < M_h/\text{M}_\odot < 10^9$), the cause behind the large scatter in the \ion{O}{i} covering fraction seems to be more complicated. 
In this case, dispersion is already present in the $\Sigma_{\ion{O}{i}} - M_h$ relation, as shown in Fig.~\ref{fig:sigma_mass}, suggesting that the enrichment and the ionisation state of the gas inside these haloes varies greatly.
In these haloes, a wide dispersion in the $\Sigma_{\ion{O}{i}} -$ SFR relation is also observed, different from high-mass haloes, where the two variables are more tightly correlated.
These results can be interpreted by considering that in these haloes star formation has begun only recently, following a bursty trend.
Moreover, most of these galaxies show SFR comparable to the minimum SFR allowed by the simulation resolution ($\sim 10^{-3}\,\rm{M}_\odot$ over a period of $10$\,Myr).
For this reason, a high fraction ($\sim 82\%$) of these haloes show null SFR, when evaluated over periods between $2-10$\,Myr, while a lower fraction $\sim 40 \%$ show null SFR on timescales of $200$\,Myr, for which the minimum SFR value allowed by the simulation resolution is lower ($\approx 5 \times 10^{-5}\,\rm{M}_\odot\,\rm{yr}^{-1}$).
Moreover, they can be more easily influenced by more massive galaxies surrounding them, which can both enrich and ionize their CGM.
However, it seems that local star formation activity also plays a role in explaining the wide scatter observed in low-mass haloes in Fig.~\ref{fig:covf_halomass}.
As we can see from the last panel in Fig.~\ref{fig:covf_sigma}, low-mass haloes with higher SFR values averaged over timescales of $200$\,Myr tend to display slightly higher covering fractions.
Star formation, in fact, favours the oxygen enrichment in the {virialised gas}, which in the central regions of haloes, where the gas density is high, is likely to be found in its neutral state.
This process would increase the \ion{O}{i} clumping factor, allowing the growth of \ion{O}{i} column density values and therefore of the \ion{O}{i} covering fraction. 
{Generally, the metal enrichment of the virialised gas following local star formation, proceeds slower than the ionising radiation field generated by the same event. 
As a result, metal enrichment in the CGM will be more tightly correlated to SFR averaged over longer timescales ($50-200$\,Myr)}.
However, the high incidence of null SFR values when averaged over short periods complicates the identification of any trend in the SFR$_{10}$ for low-mass haloes.

In conclusion, Fig.~\ref{fig:covf_sigma} highlights the role of clumpiness in the dispersion of \ion{O}{i} covering fraction values, showing how this can be related to local SFR.
In fact, SFR values measured over short timescales play a key role in determining the covering fraction of high-mass haloes, highlighting the importance of UV energetic photons generated from local star formation activity in the ionisation of the {virialised gas}.
Conversely, for low-mass haloes, higher values of SFR over longer timescales favour the presence of high-density neutral oxygen in the halo, slightly increasing the \ion{O}{i} covering fraction.
However, we do not exclude that other factors, such as the galactic environment of the halo, also have an impact, especially at lower masses.

\section{comparison with observations}
\label{sec:dndx}

Observationally, the redshift evolution of \ion{O}{i} absorbers is typically investigated using the comoving line density.
This quantity represents the density of absorbers within a comoving absorption path length, $X$, defined, following \citet{BahcallPeebles1969}, as
\begin{equation}
    \text{d}X = (1+z)^2 \frac{H_0}{H(z)} \text{d}z \, ,
\end{equation}
where $H_0$ represents the Hubble constant at the present time, and $H(z) = H_0 \sqrt{\Omega_{\rm m}(1+z)^3 +\Omega_\Lambda}$ is Friedmann's equation in a flat $\Lambda$CDM Universe with no significant contribution from radiation. 
The introduction of $\text{d}X$ ensures that the line density, $\text{d}n/\text{d}X$, of any ion is directly proportional to the product of the comoving number density and proper cross-section of the absorbers, without influences due to cosmological expansion.
To compare observational results with predictions from our simulations, we evaluate the comoving line density of neutral oxygen, $\text{d}n_{\ion{O}{i}}/\text{d}X$, considering only absorbers located {inside the virial radius} of haloes.
Our analysis began by computing the \ion{O}{i} cross-section ($\sigma$) of haloes, defined as the product between the \ion{O}{i} covering fraction and the area projected by the virial sphere ($\sigma = { f_{\rm cov}} \times \pi R_{\rm vir}^2$), which we found to follow a well-defined relationship with the halo dark matter mass, consistent with findings from previous studies \citep{Finlator2013}.
We then estimate the overall contribution of haloes within a given dark matter mass range by multiplying the typical \ion{O}{i} cross-section at each mass by the comoving number density of haloes ($\frac{\text{d}n_h}{\text{d}M_{\rm DM}}$).
This allows us to compute the absorber density per unit absorption path length and per unit dark matter mass:
\begin{equation}
\label{eq:dndXdM}
    \frac{\text{d}n}{\text{d}X\text{d}M_{\rm DM}} = \frac{c}{H_0} \sigma(M_{\rm DM}) \frac{\text{d}n_h}{\text{d}M_{\rm DM}} \, .
\end{equation}
Finally, we integrate over a specified mass range to derive $\text{d}n/\text{d}X$ of neutral oxygen for a direct comparison with observational measurements.
The following sections examine each step of this procedure in detail.

\subsection{\ion{O}{i} cross-section}

\begin{figure}
    \centering
    \includegraphics[width=\linewidth]{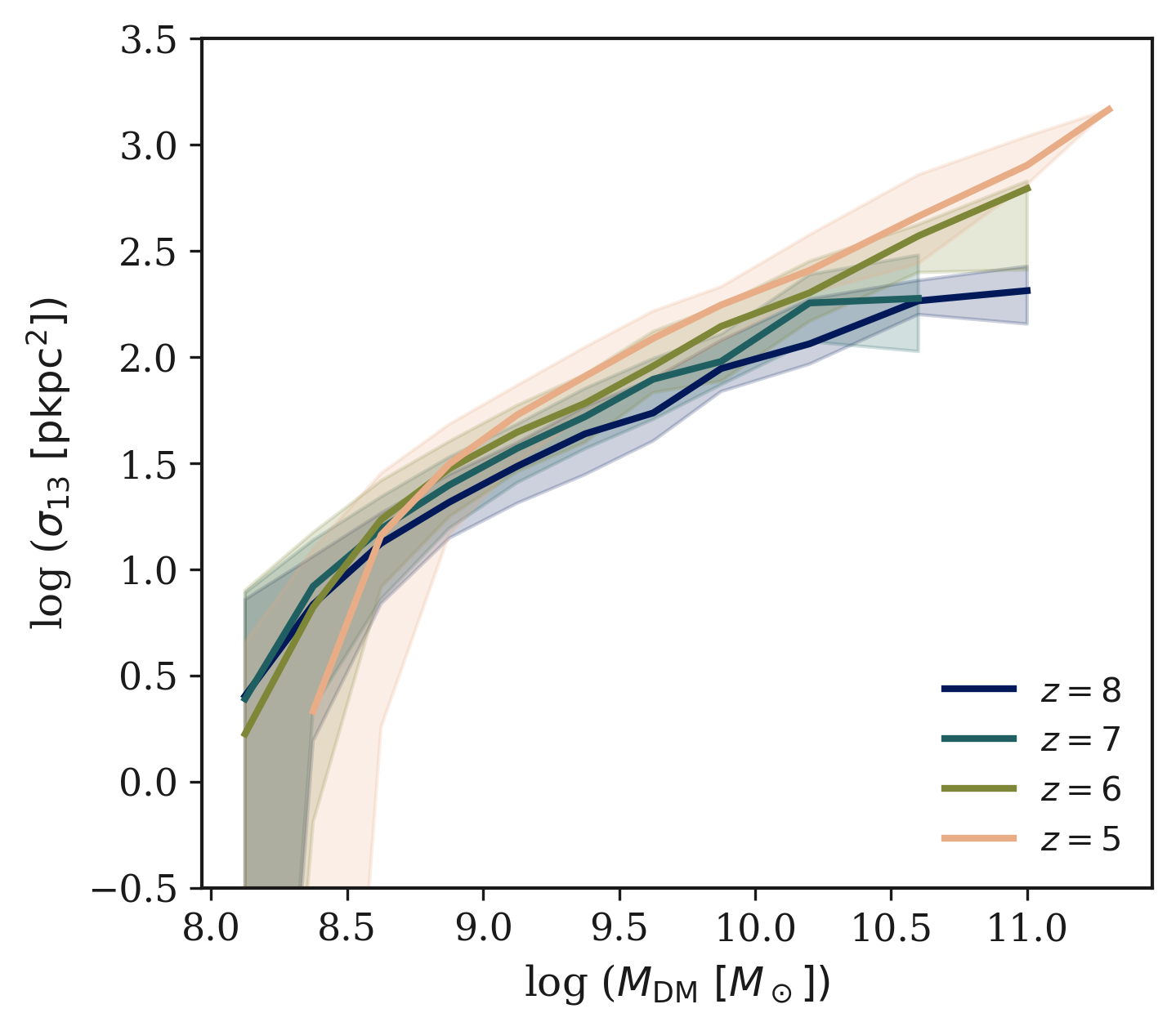}
    \caption{Neutral oxygen cross-section as a function of the halo dark matter mass and redshift. $\sigma_{13}$ is obtained as the product between the \ion{O}{i} covering fraction, computed selecting a column density threshold of $10^{13}$ cm$^{-2}$, and the area projected by the virial sphere ($\pi R_{\rm vir}^2$). The solid lines trace the median of the distribution, while the shaded regions identify the area between the $16^{\rm th}$ and $84^{\rm th}$ percentiles. A positive correlation can be observed, and despite a decrease in neutral oxygen covering fractions of high-mass haloes at lower redshift, an increase in \ion{O}{i} cross-sections caused by the increasing virial radius can be appreciated.}
    \label{fig:sigma_13}
\end{figure}

The \ion{O}{i} cross-section allows us to understand the physical size of the region {inside the virial radius} where possible absorbers reside.
The evolution of the \ion{O}{i} cross-section as a function of the halo dark matter mass and redshift is shown in Fig.~\ref{fig:sigma_13}.
Here, $\sigma_{13}$ refers to the cross-section obtained selecting covering fractions with a column density threshold of $10^{13}$ cm$^{-2}$.
However, the same trend persists when changing the threshold to $10^{14}$ cm$^{-2}$ and $10^{12}$ cm$^{-2}$, with values of $\sigma$ that can change up to $\sim 0.5$ dex in low-mass haloes, and even less at higher masses ($M_{\rm DM}\gtrsim10^9\,\text{M}_\odot$).
In the figure, a positive correlation between the cross-section and the halo dark matter mass is shown, with a higher dispersion at low masses.
The same behaviour has also been found in the previous study by \citet{Finlator2013}, although with some differences. 
In our analysis, a significant reduction in $\sigma$ at low masses occurs between $z=6$ and $z=5$, whereas in \citet{Finlator2013}, this decline begins earlier, from $z=7$.
The delay in this decrease, that in the previous analysis has been attributed to the stronger UVB that invests and ionizes low-mass haloes, highlights how in the \textsc{thesan-zoom} simulations the effects of reionization on low-mass haloes emerge slightly later, as also noted in Section~\ref{sec:evolofcovf}.

At higher dark matter masses, an opposite trend is shown, with the cross-section rising as the redshift declines.
This suggests that even if the overall \ion{O}{i} covering fraction is decreasing, as shown in Fig.~\ref{fig:covf_median_z_ev}, the growth of the virial radius becomes the dominant factor.
In \citet{Finlator2013}, this behaviour is seen only at very high masses ($\gtrsim 10^{10}\,\text{M}_\odot$), while at intermediate masses the cross section declines instead.
However, in their study, the \ion{O}{i} covering fraction of intermediate-mass haloes ($10^8 < M_{\rm DM}/\text{M}_\odot < 10^9$) shows a more significant drop with decreasing redshift, which likely drives the observed trend in $\sigma$.
Our results suggest that in the \textsc{thesan-zoom} simulations the neutral areas close to star-forming regions expand while approaching lower redshifts, while the gas begins to be ionized further away from the source and at lower overdensities when compared to earlier studies.
These differences likely arise because of multiple factors, such as the later reionization history and the higher gas resolution in the \textsc{thesan-zoom} simulation, but also the more bursty star formation, which could help drive the gas to larger distances from the halo.
Overall, in our simulations, the effects of reionization seem to be directly observed in the \ion{O}{i} cross-sections of low-mass haloes ($M_{\rm DM} < 10^9 \ \rm{M}_\odot$), where the \textsc{thesan-zoom} simulations have already showed how reionization is able to unbind the gas, lowering the baryonic fraction of these haloes from $z=7$ to $z=5$ \citep[][]{Zier2025}. 
However, due to the large variance, observational tests would require very large samples to confirm our model.
On the other hand, haloes with dark matter mass above $10^9\,\text{M}_\odot$ show larger areas of high-density neutral gas as we move to lower redshift.
Yet, this result, together with the decrease displayed by the \ion{O}{i} covering fraction, suggests that even if the neutral gas occupies larger regions around high-mass haloes as reionization progresses, it is confined to higher overdensities.

Following \citet{Finlator2013}, at each redshift, we modelled the median values of $\sigma$ with a broken power law.
Each part of the curve follows the relation $\log_{10} (\sigma/{\rm pkpc}^2)=m \log_{10}(M_{\rm DM}/\text{M}_\odot) +q$, with best-fit parameters that shift from $(m_l, q_l)$ to $(m_h, q_h)$ at a value of the dark matter mass $M_c$, which, by imposing continuity, has to be $\log_{10} (M_c/\text{M}_\odot) = (q_h - q_l)/(m_l - m_h)$.
The best-fit parameters are reported in Appendix~\ref{app:best-fitsigma}, and they are used to determine the function $\sigma (M_{\rm DM})$ to use in the expression outlined by Eq.~(\ref{eq:dndXdM}).

\subsection{\ion{O}{i} comoving line density}

\begin{figure}
    \centering
    \includegraphics[width=\linewidth]{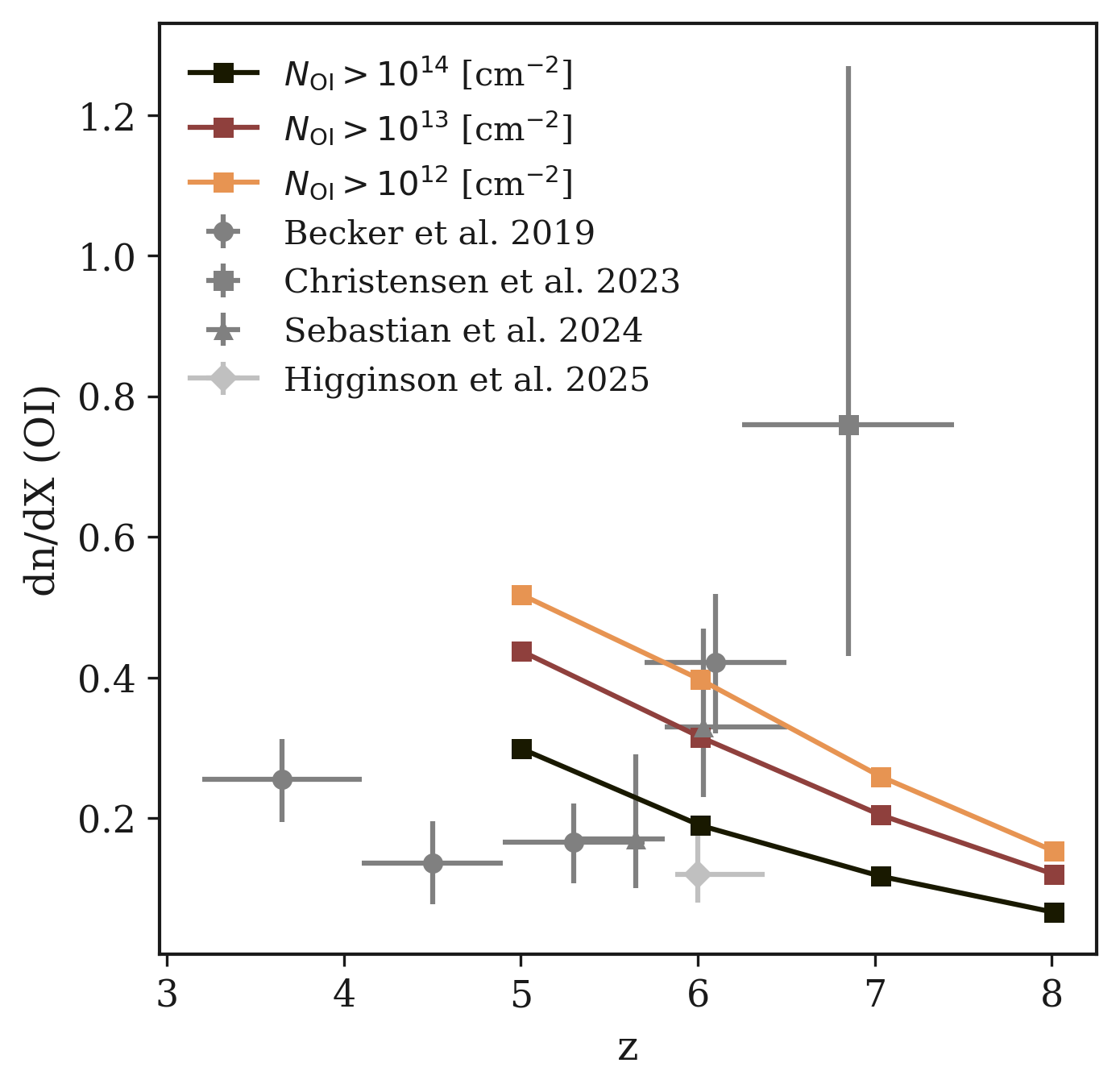}
    \caption{Redshift evolution of the \ion{O}{i} absorbers comoving line density. The coloured squares represent our results when considering only possible absorbers in the {virial radius} of galaxies with dark matter mass above $10^8\,\text{M}_\odot$. Darker (lighter) points mark absorber densities when selecting higher (lower) column density thresholds ($10^{14}$ cm$^{-2}$, $10^{13}$ cm$^{-2}$, $10^{12}$ cm$^{-2}$).
    The grey data points identify observational results from \citet{Becker2019} (circles, EW>0.05 \AA, $N_{\ion{O}{i}}\gtrsim 10^{13.7}$ cm$^{-2}$), \citet{Christensen2023} (squares, EW>0.05 \AA, $N_{\ion{O}{i}}\gtrsim 10^{13.7}$ cm$^{-2}$) and \citet{Sebastian2024} (triangles, EW>0.03 \AA, $N_{\ion{O}{i}}\gtrsim 10^{13.9}$ cm$^{-2}$) and \citet{Higginson2025} (light gray diamond, EW>0.05 \AA, $N_{\ion{O}{i}}\gtrsim 10^{13.7}$ cm$^{-2}$). The apparent discrepancy is likely due to the presence of possible absorbers outside the {virial radius}, as well as to differences in the likelihood of sampling the different regions of the gas with observations.}
    \label{fig:dndX}
\end{figure}

Since the \textsc{thesan-zoom} simulations follow small volumes at high resolution, they are not ideally suited for the evaluation of halo mass functions.
For this reason, in order to compute the density of absorbers following Eq.~(\ref{eq:dndXdM}), we used the python package \texttt{hmf} \citep[][]{Murray_hmf_2013} to estimate $\text{d}n_h/\text{d}M_{\rm DM}$, adopting the cosmological parameters used in the simulations.
We numerically integrated the expression in Eq.~(\ref{eq:dndXdM}), selecting a minimum dark matter mass of $10^8\,\text{M}_\odot$, which translates to slightly higher values at $z=5$, since the median $\sigma$ drops to zero at $M_{\rm DM} \lesssim 1.8 \times 10^8\,\text{M}_\odot$.
The integration is operated up to a maximum of $10^{12}\,\text{M}_\odot$, after noticing that the contribution from more massive haloes would increase the result of less than $1\%$, because of their low number density.
In this way, we obtained for each redshift a value of $\text{d}n/\text{d}X$ computed using different column density thresholds.
The results are reported in Table~\ref{tab:dndX} and represented as coloured squares in Fig.~\ref{fig:dndX}, where they are compared to observations based on blind quasar absorption line surveys \citep[dark gray;][]{Becker2019, Christensen2023,  Sebastian2024} and on association of galaxies and absorbers following a galaxy-centric approach \citep[light gray;][]{Higginson2025}.

Some features can make the comparison between observational data challenging.
First, not all absorbers have the same probability of being detected, since those with a higher EW tend to emerge more easily.
For this reason, observational studies correct the number of observed absorbers with a suitable completeness function, which is unique for each study.
Moreover, a selection on absorbers needs to be operated when computing the comoving line density. 
Usually, only absorbers with an EW above a threshold value and that do not reside in quasar proximity zones are considered.
However, different studies tend to use different criteria, and drawing direct comparisons between results becomes more challenging.
The observational data reported in Fig.~\ref{fig:dndX} are computed considering absorbers with EW$>0.05$ \AA \ \citep[][]{Becker2019, Christensen2023, Higginson2025} or EW$>0.03$ \AA \ \citep[][]{Sebastian2024}.
Assuming the curve of growth computed by \citet{Wang2020}, these values should correspond to $N_{\ion{O}{i}}$ of $\approx 10^{13.9}$ cm$^{-2}$ and $\approx 10^{13.7}$ cm$^{-2}$, when assuming the Voigt $b$ parameter to be within $10 - 20$\,km\,s$^{-1}$.
Moreover, the proximity regions considered in these studies differ, and as shown in \citet{Sebastian2024}, at $z\gtrsim6$, a difference of $0.2$ in $\text{d}n/\text{d}X$ can arise when switching from proximity regions of $10\,000$\,km\,s$^{-1}$ to $3000$\,km\,s$^{-1}$.

Despite the different absorber selection, observational results based on blind quasar absorption-line surveys seem to consistently reproduce a rise in $\text{d}n/\text{d}X$ of \ion{O}{i} absorbers from $z=5$ to $z=8$.
Predictions from our simulations represented by the black line in Fig.~\ref{fig:dndX} would be expected to follow the same trend traced by observational points if all the absorbers were inside the {virial radius} of galaxies with DM mass above $10^8\,\text{M}_\odot$.
However, our results fail to reproduce this rising trend, suggesting a decline at higher redshifts, instead.
The same behaviour is present also when considering weaker absorbers, with column densities of $10^{13}$ cm$^{-2}$ or below, that might be observable in the future also thanks to ArmazoNes high Dispersion Echelle Spectrograph \citep[ANDES, ][]{D'Odorico2023}.
This decline toward high redshift has also been found in the previous work of \citet{Finlator2013}, where a similar approach has been followed.
However, this method does not account for absorbers that may lie beyond the virial radius of the haloes.

Following a galaxy-centric method, the recent observational results of \citet{Higginson2025} also show the high incidence of neutral oxygen absorbers outside the virial radius of high-redshift galaxies. 
In their work, the \ion{O}{i} covering fraction around [\ion{O}{iii}]-emitting galaxies is used to evaluate the line density of neutral oxygen absorbers at $z=6$, with an approach similar to the one used in our work.
Their result, shown as a light gray diamond in Fig.~\ref{fig:dndX}, is lower than what blind quasar surveys suggest, implying that most of the neutral oxygen ($\approx 65\%$) should reside at larger distances from massive detectable galaxies ($M_\star \gtrsim 10^{7.5}\,\rm{M}_\odot$) at $z=6$, close to low-mass sources or in dense neutral IGM pockets that survived reionization.
Based on both our results and these recent observations, we investigate in the next section the role of neutral oxygen outside the {virial radius} of haloes considered in our study.

\begin{table}
    \centering
    \addtolength{\tabcolsep}{4pt}
    \begin{tabular}{cccc}
        \hline
        Redshift & $(\text{d}n_{\ion{O}{i}}/\text{d}X)_{14}$ & $(\text{d}n_{\ion{O}{i}}/\text{d}X)_{13}$ & $(\text{d}n_{\ion{O}{i}}/\text{d}X)_{12}$ \\
        \hline
        \hline
        5 & 0.3019 & 0.4375 & 0.5237\\
        6 & 0.1914 & 0.3188 & 0.4021\\
        7 & 0.1184 & 0.2070 & 0.2629\\
        8 & 0.0666 & 0.1212 & 0.1555\\
        \hline
    \end{tabular}
    \caption{Comoving line density results represented in Fig.~\ref{fig:dndX}.
    At each redshift we are reporting the values of $\text{d}n/\text{d}X$ for neutral oxygen absorbers within the virial radius of haloes, when selecting column density thresholds of $10^{14}$ cm$^{-2}$, $10^{13}$ cm$^{-2}$ and $10^{12}$ cm$^{-2}$.}
    \label{tab:dndX}
\end{table}

\subsection{\ion{O}{i} outside the virial radius}
\label{sec:OI_outside}

\begin{figure}
    \centering
    \includegraphics[width=\linewidth]{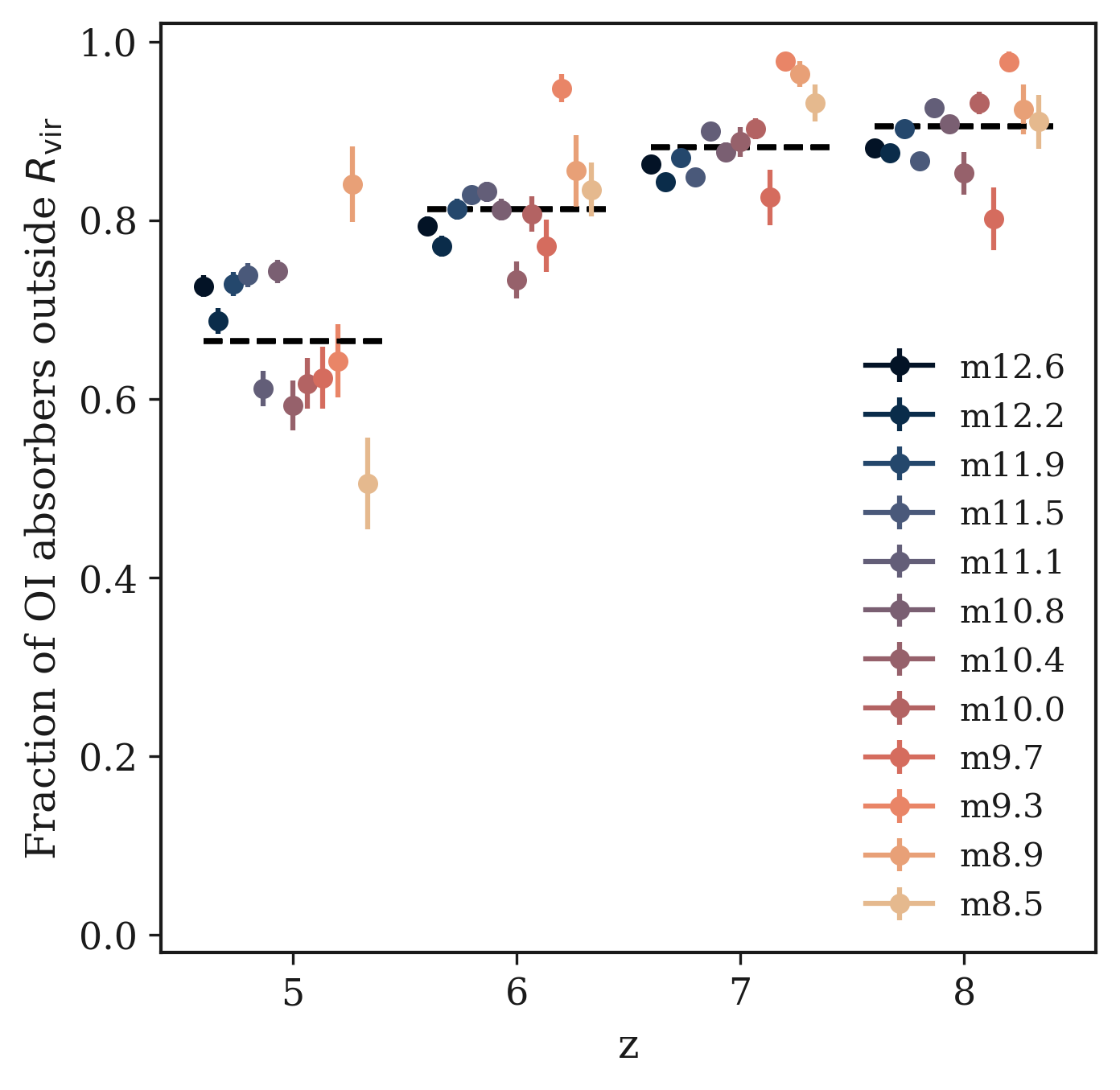}
    \caption{Fraction of \ion{O}{i} absorbers with column density $N_{\ion{O}{i}}> 10^{13}$ cm$^{-2}$ outside the virial radius of haloes with mass $M_{\rm DM} > 10^8\,\text{M}_\odot$. The coloured points represent the different zoom-in regions analysed which have been slightly shifted on the $x$-axis for visual purposes. While the dashed black line marks the median value obtained when considering all the regions at a given redshift. The error bars mark the uncertainties obtained from binomial statistics. The number of unaccounted absorbers increases at higher redshift.}
    \label{fig:OI_outside_13}  
\end{figure}

To estimate the contribution of absorbers located outside the {virial radius of haloes} to the overall \ion{O}{i} line density, we focused in each zoom-in region on all the gas cells reproduced at high resolution.
Considering these cells, we projected the $N_{\ion{O}{i}}$ values on a grid, along a single direction.
The cell size of the grid varies between $0.5 - 1.25$\,pkpc depending on the dimension of the high-resolution region. 
We verified that varying the grid resolution within this range does not significantly affect the overall column density distribution of absorbers.
In each region, we then selected a number $N$ of grid cells hosting potential absorbers, i.e., cells with the computed $N_{\ion{O}{i}}$ value above a specified threshold. 
The number $N$ differs across zoom-in regions and is determined by multiplying the fraction of the projected area occupied by potential absorbers by $10^4$.
Thus, $N$ represents the number of sightlines intersecting absorbers that we would expect if we randomly drew $10^4$ lines of sight through the zoom-in region.
Along each sightline, we evaluated $N_{\ion{O}{i}}$ in segments of $5$\,kpc and identified those with column density above the threshold.
If multiple segments along the line of sight satisfy this condition, we consider them as a single absorber if they extend for less than $50$\,km\,s$^{-1}$, which at the analysed redshifts corresponds to $80 - 120$\,pkpc.
In practice, even in the densest regions, the extent of individual absorbers along sightlines does not exceed $\approx 65$\,pkpc.
Such absorbers, however, can only be found in very dense zoom-in areas and are more common at lower redshifts.
For each absorber, we calculated the median position along the sightline and determined whether it lies outside the virial radius of haloes with $M_{\rm DM}>10^8\,\text{M}_\odot$.

The fraction of \ion{O}{i} absorbers with $N_{\ion{O}{i}}>10^{13}$ cm$^{-2}$ outside the {virial radius} of these haloes are represented in Fig.~\ref{fig:OI_outside_13}.
Each point identifies the result obtained in a given high-resolution region of the simulations, with error bars indicating uncertainties derived from binomial statistics.
The dashed black line marks the median value among all the zoom-in regions at each redshift.
This value increases at higher redshifts, highlighting how neglecting the absorbers located outside the {virial radius} might explain the observed decline in the line density represented in Fig.~\ref{fig:dndX}.
A similar result has been presented in \citet{Doughty2022}, where the authors show that \ion{O}{i} at low gas overdensity, i.e., outside the virial radius, is more common at higher redshift, as our data suggest.
Observationally, \citet{Higginson2025} has detected neutral oxygen absorbers as far as $8\,R_{\rm vir}$ from [\ion{O}{iii}]-emitting galaxies, implying that a large amount of neutral oxygen is not bound to detectable galaxies during reionization.

The trend does not change even when considering a higher column density threshold of $10^{14}$ cm$^{-2}$, though the corresponding values are on average $20\%$ lower.
Due to the extended nature of absorbers, pinpointing their exact location along the line of sight can be challenging.
{To perform the analysis introduced in this section, we considered the median position of the absorber along the sightline. 
However, even if the median position lies outside the virial radius, the absorber may still extend within it}.
The reverse is also true: absorbers with median position inside the {virial radius} of galaxies might extend outside of it.
Nonetheless, we noticed that, with the applied technique, the impact of the chosen segment size of $5$\,kpc over which we evaluated the column densities along the line of sight is minimal. 
In fact, even when using a finer grid of size $1$\,pkpc, the median value of absorbers residing outside the {virial radius} only presents a minor decrease of $\approx 3\%$. 

\begin{figure*}
    \centering
    \includegraphics[width=\linewidth]{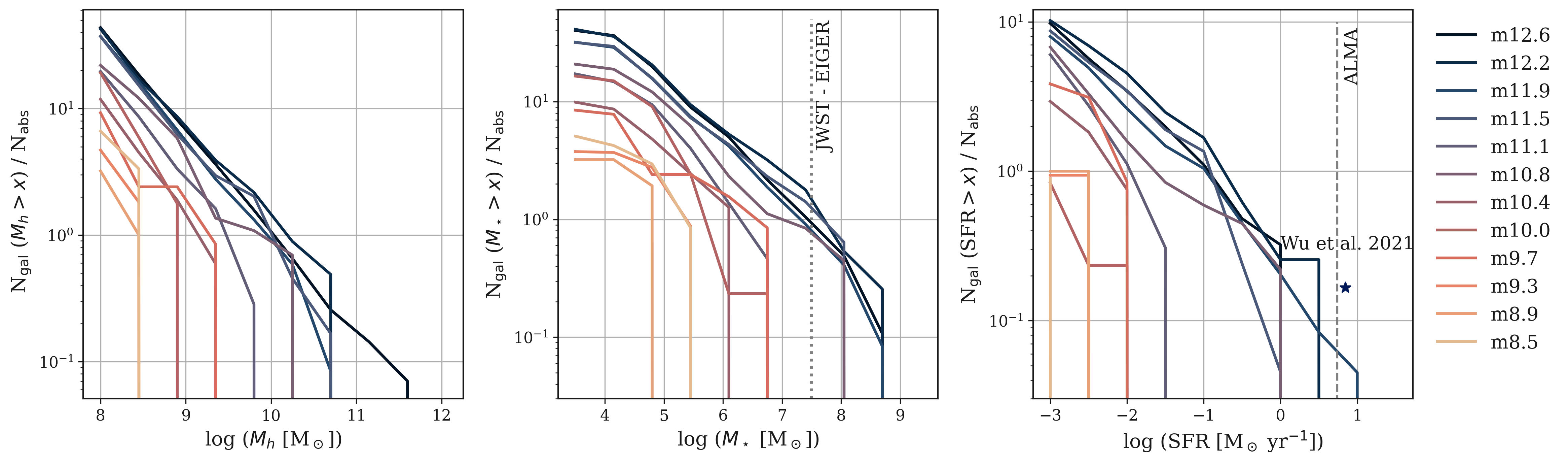}
    \caption{Galactic environment around \ion{O}{i} absorbers with column density above $10^{14}$ cm$^{-2}$ at $z=6$. The plots show the number of predicted galaxies per absorber as a function of galaxy properties as halo mass, stellar mass and SFR when selecting a volume around the absorbers comparable with ALMA observations in \citet{Wu2021,Wu2023}. Coloured lines represent results obtained from different zoom-in regions, while the detection in \citet{Wu2021} is indicated in the third panel by the black star. The grey vertical lines mark the detection thresholds of \textit{JWST} EIGER survey \citep[][]{Higginson2025} and ALMA \citep[][]{Wu2021}, in the second and third panels respectively. Low-mass and low-SF galaxies are more common to observe around absorbers, but current observations are still only sensitive to more massive and bright galaxies. }
    \label{fig:wucomp}
\end{figure*}

\subsection{The galactic environment around absorbers}
\label{sec:galenvironment}

Connecting absorbers to galaxies represents one of the major challenges for both theoretical simulations and observational studies.
At high redshift, \citet{Wu2021, Wu2023} found a galaxy with an estimated SFR of $7\,\text{M}_\odot$ yr$^{-1}$ located at $\approx 20$\,kpc from a \ion{O}{i} absorber with $N_{\ion{O}{i}} = 10^{14.2}$ cm$^{-2}$ at $z = 5.978$.
Earlier simulations \citep[][]{Finlator2020} showed how in the surroundings of each absorber approximately $10^{-2.5}$ galaxies similar to this are expected to be found.
However, the detection of one such galaxy across only six quasar sightlines suggests that the actual observational probability might be higher.
With the advantage of the improved spatial resolution in the \textsc{thesan-zoom} simulations, we revisited this analysis to investigate the probability of detecting similar galaxies around \ion{O}{i} absorbers.

We identified \ion{O}{i} absorbers in each zoom-in region following the method explained in Section~\ref{sec:OI_outside}.
For each absorber, we explored the properties of galaxies within an area comparable to that probed by ALMA observations of \citet{Wu2023}.
Specifically, we selected all galaxies with an impact parameter less than $50$\,pkpc and a velocity offset within $200$\,km\,s$^{-1}$ from the selected absorbers.
The halo mass, stellar mass and SFR of these galaxies are represented in Fig.~\ref{fig:wucomp}, where, for comparison, only absorbers with $N_{\ion{O}{i}}>10^{14}$ cm$^{-2}$ have been considered.
Each panel displays the number of galaxies per absorber exceeding the values of halo mass, stellar mass or SFR reported on the $x$-axis.

The solid lines, which show results for different zoom-in regions, reveal an overall consistent trend: lower mass and low star-forming haloes are more numerous.
Moreover, absorbers in highly-dense zoom-in regions are more likely to be surrounded by a higher number of galaxies, as indicated by the higher values displayed by darker lines.
The maximum halo mass present is also correlated to the density of the zoom-in region and, as a result, massive and highly star-forming galaxies are absent in low-density regions (light coloured lines).
Similar behaviours are observed even when considering weaker absorbers, which can also be found in more distant regions from galaxies, and therefore display slightly lower numbers of associated sources.
The black star in the third panel marks the observational results from \citet{Wu2021, Wu2023}, while their S/N threshold of $5.5\,\text{M}_\odot$ yr$^{-1}$ is identified by the gray dashed line. 
In our simulations, galaxies comparable to the observed ones could be found only in {high-density zoom-in regions, i.e. those that at $z=3$ form haloes more massive than $10^{12}\,M_\odot$, where the predicted number of such sources around absorbers is $\approx 10^{-1.5}$. 
This means that $1$ out of $30$ absorbers crossing a high-density region would show a galaxy similar to the one observed by ALMA. 
However, for blind quasar surveys, we would expect this to be an upper limit, since also the probability of observing an absorber in one of these regions needs to be accounted.}

Recent \textit{JWST} observations highlighted how only $\approx 35\%$ of neutral oxygen absorbers are found in the vicinity of detectable [\ion{O}{iii}]-emitting galaxies \citep[][]{Higginson2025}, implying that the majority of neutral oxygen likely resides close to lower-mass galaxies below the EIGER survey detection threshold.
This detection threshold of $M_\star \gtrsim 10^{7.5}\,\rm{M}_\odot$, as indicated by the dotted line in the second panel of Fig.~\ref{fig:wucomp}, highlights how high-redshift observations are able to capture mainly high-mass galaxies that reside in overdensities, missing the more frequent low-mass sources. 
Moreover, the observed absorbers and galaxies are usually separated by large distances ($\approx 100$ pkpc), which, according to our simulation data, suggests the possible presence of lower-mass undetected sources located closer to the observed absorbers, especially in high-density regions.

In conclusion, with the \textsc{thesan-zoom} simulations we demonstrated how low-mass and low star-forming galaxies are more likely to be found around absorbers, as also previous simulations and observations pointed out \citep[][]{Keating2014, Finlator2020, Higginson2025}, with absorbers in highly dense regions being surrounded by an overall larger number of sources.
Moreover, the simulations are able to reproduce observations similar to the one in \citet{Wu2023}, although such an event is considered to be rare.
In fact, we estimate that $1$ in $\approx 30$ absorbers in high-density regions should be surrounded by a galaxy detectable with ALMA observations similar to \citet{Wu2023}.
Assuming this is the case, the probability of detecting one such galaxy among six quasar absorption lines in a high-density environment would be $20\%$.

\section{Conclusions}
\label{sec:conclusions}

In this work, we used the \textsc{thesan-zoom} simulations to investigate the connection between \ion{O}{i} absorbers and galaxies in the redshift range $z=5 - 8$. 
The simulations track the abundance of five chemical species, including oxygen, as well as the various ionisation states of hydrogen and helium, which are computed on-the-fly, fully coupled to the radiation field.
Assuming that the neutral oxygen fraction mirrors that of neutral hydrogen (Eq.~\ref{eq:HOeq}), we followed neutral oxygen throughout $13$ zoom-in regions tracing different galaxy environments.
We investigated the properties of potential \ion{O}{i} absorbers, starting from haloes with mass $> 10^8\,\text{M}_\odot$ and their virialised gas.
In the following, we summarise the main findings.
\begin{enumerate}
    \item Observable neutral oxygen absorbers ($N_{\ion{O}{i}} \gtrsim 10^{13}$ cm$^{-2}$) are associated with LLSs or DLAs, as also previous works suggested. 
    However, the precise relation between \ion{O}{i} and \ion{H}{i} column density is complicated by the inhomogeneity of chemical enrichment.
    In fact, although we notice in Fig.~\ref{fig:rad_prof_cd_10} how \ion{H}{i} and \ion{O}{i} average radial profiles follow the same evolution, we show that converting $N_{\ion{O}{i}}$ of individual absorbers into $N_{\rm HI}$ can introduce uncertainties of up to three orders of magnitude (Fig.~\ref{fig:OI_to_HI}).
    Moreover, we demonstrate how information on the overall metallicity of gas in absorption can break down the degeneracy between neutral oxygen column density and the fraction of neutral hydrogen, essential to characterise reionization history.

    \item We find a decreasing neutral oxygen covering fraction in massive haloes toward lower redshifts, which results from gas within the virial radius becoming increasingly ionised with cosmic time. 
    We think that the local ionising field together with the overall reionization process are the primary drivers of this transformation, although we do not rule out minor contributions due to SN feedback heating the gas.
    However, a more detailed analysis is required to quantify the relative contribution of each component, which will be the focus of future work.

    \item We find that if neutral oxygen was homogeneously distributed {inside the virial radius}, a positive correlation between the \ion{O}{i} covering fraction and halo mass is expected. However, our data display a dispersion about twice as large as that reported in earlier similar works, making it difficult to clearly identify such a correlation.
    The primary source of this scatter, especially in higher-mass haloes ($M_h>10^9\,\rm{M}_\odot$), is the clumpiness of \ion{O}{i}.
    The clumping factor is positively correlated with SFR inside the halo, especially at higher masses. As a result, part of the dispersion can be explained by SFR. In fact, at $M_h\gtrsim10^9\,\text{M}_\odot$, a higher SFR on small timescales has the effect of ionising the gas in the CGM, decreasing the \ion{O}{i} covering fraction.
    In low-mass haloes, however, SFR has the main effect of producing a sufficient amount of oxygen in order to have detectable absorbers, and as a consequence, haloes with higher SFR on longer timescales show on average higher \ion{O}{i} covering fractions. We do not exclude that additional factors, as the effects of local environment, can play an important role alongside SFR in determining the covering fraction, especially for low mass haloes, where the dispersion in \ion{O}{i} covering fraction values is larger.

    \item To consistently reproduce the line density of absorbers and their evolution that are comparable to observations, it is necessary to consider potential absorbers outside the virial radius of galaxies. 
    We find that especially at high redshift ($z=8$), the median position of $\sim 90\%$ of the absorbers is found outside the {virial radius}. Neglecting these absorbers would produce a decreasing incidence rate with increasing redshift, contrary to observational findings.

    \item Detections of galaxies surrounding absorbers \citep{Wu2023} are expected only in high-density regions, with a probability of {at least} a factor $5$ lower than that suggested by observations.
    Overall, as highlighted in Fig.~\ref{fig:wucomp}, absorbers are more likely to be surrounded by more numerous low-mass and low star-forming galaxies that cannot be detected by current instruments, explaining the lack of galaxy counterparts highlighted by most observations.
\end{enumerate}

Future observational studies will contribute to the building of a more extensive catalogue of absorbers and their host galaxies, potentially revealing similarities and discrepancies with theoretical models. 
In fact, multiple \textit{JWST} collaborations \citep[e.g.,][]{Kashino2023, Zou2024} will continue to focus on the interplay between galaxies and gas at high redshift, allowing for a more complete census of sources near detected absorbers. 
In the longer term, the future spectrographs HARMONI \citep[High Angular Resolution Monolithic Optical and Near-infrared Integral field spectrograph][]{harmoni2021} and ANDES on the ELT will be essential to gain high-resolution data of absorbers in the CGM.
Moreover, cross-correlation of data from SKA \citep[Square Kilometer Array,][]{ska2009} and \textit{JWST} will also be essential for a more complete view of the interaction between gas and galaxies.

Overall, while the zoom-in simulations allowed us to reach a high resolution, necessary to resolve the small and intricate gas structures around galaxies, precise comparisons with observations remain challenging. 
In future works, the modelling of neutral oxygen could be refined by combining the radiation field information with chemical abundances in post-processing. 
This approach would not only allow us to track additional ionisation states of oxygen, but it will also enable the study of other metal absorbers, such as \ion{C}{ii}, \ion{Mg}{ii}, \ion{C}{iv} and \ion{Si}{iv}, offering a more comprehensive view of the overall absorber population predicted by the \textsc{thesan-zoom} simulations.
Moreover, integrating prescriptions for line-emitting galaxies in the simulations could provide a more straightforward method to compare theoretical models to observations.
Techniques to reproduce the population of Ly$\alpha$ and metal line emitters in simulations, such as the Cosmic Ly$\alpha$ Transfer code \citep[COLT;][]{Smith2015, Smith2019, Smith2022MW}, which has been successfully used also at low redshift \citep[][]{Tacchella2022, McClymont2024}, will be applied in the future to the \textsc{thesan-zoom} simulations, including extracting detailed absorption spectra to complement our present study.
In this way, we could potentially select all the galaxies that are more likely to be detected through slitless spectroscopy thanks to their powerful emission lines, providing more accurate predictions for future observations.
Furthermore, comparisons with other high-resolution simulations such as FIREbox \citep[][]{Feldmann2023} are essential for identifying strengths and limitations of each model, representing an important step toward improving our understanding and modelling of gas in the Universe.

\section*{Acknowledgements}
{The authors sincerely thank the anonymous reviewer for the constructive and encouraging feedback.}
We gratefully acknowledge the Gauss Centre for Supercomputing e.V. (\href{www.gauss-centre.eu}{www.gauss-centre.eu}) for funding
this project by providing computing time on the GCS Supercomputer
SuperMUC-NG at Leibniz Supercomputing Center (\href{www.lrz.de}{www.lrz.de}),
under project pn29we.
GP thanks Sankalan Bhattacharyya for the very useful discussion on the nature of covering fractions and all the participants of the Galaxy-IGM IFPU workshop for the interesting interactions. LK acknowledges the support of a Royal Society University Research Fellowship (grant number URF$\backslash$R1$\backslash$251793). RK acknowledges support of the Natural Sciences and Engineering Research Council of Canada (NSERC) through a Discovery Grant and a Discovery Launch Supplement (funding reference numbers RGPIN-2024-06222 and DGECR-2024-00144) and York University's Global Research Excellence Initiative.
XS acknowledges support from the NASA theory grant JWST-AR-04814.
WM thanks the Science and Technology Facilities Council (STFC) Center for Doctoral Training (CDT) in Data Intensive Science at the University of Cambridge (STFC grant number 2742968) for a PhD studentship.

%%%%%%%%%%%%%%%%%%%%%%%%%%%%%%%%%%%%%%%%%%%%%%%%%%
\section*{Data Availability}

All simulation data, including snapshots, group, subhalo catalogues and merger trees will be made publicly available in the near
future via \href{www.thesan-project.com}{www.thesan-project.com}. Before the public data release, data underlying this paper will be shared on reasonable request to the corresponding author.

%\begin{figure*}
    %\centering
    %\includegraphics[width=0.9\linewidth]{figures/OI_covf_sfr.png}
    %\caption{Caption}
    %\label{fig:enter-label}
%\end{figure*}



\bibliographystyle{mnras}
\bibliography{paper} % if your bibtex file is called example.bib

%\begin{thebibliography}{99}
%\end{thebibliography}

%%%%%%%%%%%%%%%%%%%%%%%%%%%%%%%%%%%%%%%%%%%%%%%%%%

%%%%%%%%%%%%%%%%% APPENDICES %%%%%%%%%%%%%%%%%%%%%

\appendix

\section{Best-fit parameters for the $x_{\ion{H}{i}} - N_{\ion{O}{i}}$ relation}
\label{app:best-fit_HI_NOI}

\begin{table}
    \centering
    \begin{tabular}{cccc}
        \hline
          metallicity & $a$ & $b$ & $c$ \\
        \hline
        $Z/\text{Z}_\odot < -2.75$ & 0.9575 & 1.4196 & 12.4783 \\
        $-2.75 < Z/\text{Z}_\odot < -2.25$ & 0.9437 & 1.6361 & 13.2554 \\
        $-2.25 < Z/\text{Z}_\odot < -1.75$ & 0.9326 & 1.6743 & 13.7742 \\
        $-1.75 < Z/\text{Z}_\odot < -1.25$ & 0.9177 & 1.8546 & 14.2778 \\
        $-1.25 < Z/\text{Z}_\odot < 0.00$ & 0.9132 & 1.9121 & 14.8466 \\
        \hline
    \end{tabular}
    \caption{Best-fit parameters of the sigmoid function used to describe at different metallicities the relation between the fraction of neutral gas and the neutral oxygen column density.}
    \label{tab:x_HI_N_OI}
\end{table}

We fit the median values of the fraction of neutral hydrogen as a function of $N_{\ion{O}{i}}$ in different metallicity bins with a sigmoid function defined as
\begin{equation}
    x_{\ion{H}{i}} = \frac{a}{1 + e^{-b(x-c)}}
\end{equation}
The best-fit parameters defining the dashed lines in Fig.\ref{fig:f_HI_metal} are reported in Table~\ref{tab:x_HI_N_OI}.

\section{Covering fraction and surface density in the homogenous case}
\label{app:covfrac_theo}

\begin{figure}
    \centering
    \includegraphics[width =0.7\columnwidth]{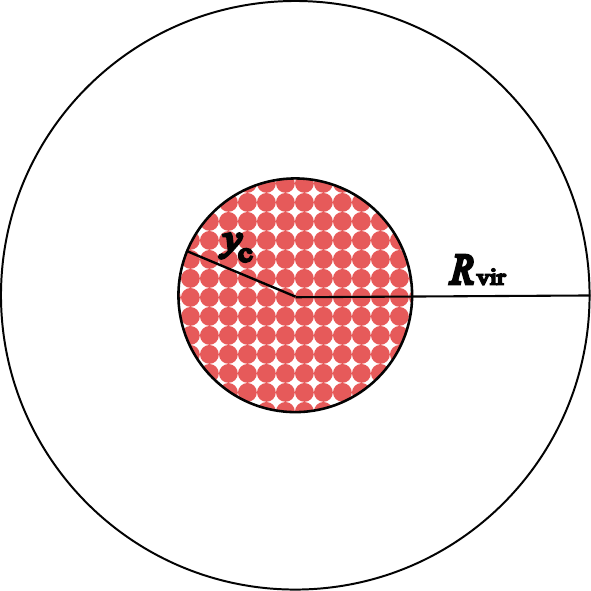}
    \caption{Representation of the covering fraction for a homogeneously distributed element. The larger circle of radius $R_{\rm vir}$ is the projection of the virial sphere of a halo in which an element $x$ is homogeneously distributed. In such a sphere, column density values only depend on the projected distance between the centre and the line of sight. In this case, we suppose that the selected threshold column density ($N_{\rm th}$) is the value that can be measured along lines of sight with distance $y_c$ from the centre. Therefore, the covering fraction is obtained as the ratio between the red-dotted area and the area of the larger circle. }
    \label{fig:covfrac_draw}
\end{figure}

{In the virialised gas} of a halo where an element or ion, $x$, is homogeneously distributed, the covering fraction of this element can be determined using the virial radius of the halo ($R_{\rm vir}$) and the total mass of $x$ contained within it ($M_x$).
Since the element is homogeneously distributed inside a sphere, the values of column density at a given location change only according to the length of the line of sight.
This length varies as a function of the projected distance, $y$, between the centre of the sphere and the line of sight.
Column density values of the element $x$, at a given projected distance $y$ from the centre, are obtained as
\begin{equation}
\label{eq:N_x}
    N_{x} (y) = 2n_{x} \sqrt{R_{\rm vir}^2 - y^2} \, ,
\end{equation}
where $2 \sqrt{R_{\rm vir}^2 - y^2}$ is the length of the line of sight, and $n_{x}$ the number density of $x$, that can be expressed as
\begin{equation}
\label{eq:nx}
    n_{x} = \frac{M_{x}}{\frac{4}{3} \pi R_{\rm vir}^3 m_{x}} \, ,
\end{equation}
with $m_x$ being the atomic mass of element $x$.
The highest values of column density are found close to the centre, where lines of sight are longer. 
Therefore, the covering fraction can be defined as the ratio between the area of the circle with radius $y_{c}$, inside which column densities are higher than the selected threshold ($N_{\rm th}$) over the total projected area of the sphere:
\begin{equation}
\label{eq:covf}
    f_{\rm cov} = \frac{y_c^2}{R_{\rm vir}^2} \, .
\end{equation}
A schematic representation of this is provided in Fig.~\ref{fig:covfrac_draw}.

The distance $y_{c}$ can be expressed as a function of the threshold column density and the number density of the element, $n_x$, using Eq.~(\ref{eq:N_x}). 
However, if the density is not high enough and the virial radius is small, even the column density along the diameter of the sphere could be lower than the threshold.
In that case, $y_c$ is set to zero:
\begin{equation}
y_c = \left\{\begin{split}
&R_{\rm vir}^2 - \left( \frac{N_{\rm th}}{2 n_{x}}\right)^2  \ \ \ \ \ \ \ \ & {\rm if}  \ \ \frac{N_{\rm th}}{2 n_{x}} \leq R_{\rm vir} \\
&0 &{\rm otherwise} \ \ \ \ \ \\ 
\end{split}\right. \, .
\end{equation}

Substituting the expression found for $y_c$ in Eq.~(\ref{eq:covf}), expressing $n_x$ using Eq.~(\ref{eq:nx}) and introducing the surface density as $\Sigma_x = M_x/(\pi R_{\rm vir}^2)$, the covering fraction can be written as

\begin{equation}
\label{eq:covf_logsigma}
    f_{\rm cov} = \left\{\begin{split}
    &1 - \left(\frac{k}{10^{\log \Sigma}}\right)^2 \ \ \ \ \ \ \ \ & {\rm if}  \ \ \Sigma\geq k \\
    &0 &{\rm otherwise}\\ 
\end{split}\right. \, ,
\end{equation}
where $k = 2/3 N_{\rm th} m_x$ is a constant value. 

\begin{figure}
    \centering
    \includegraphics[width=\columnwidth]{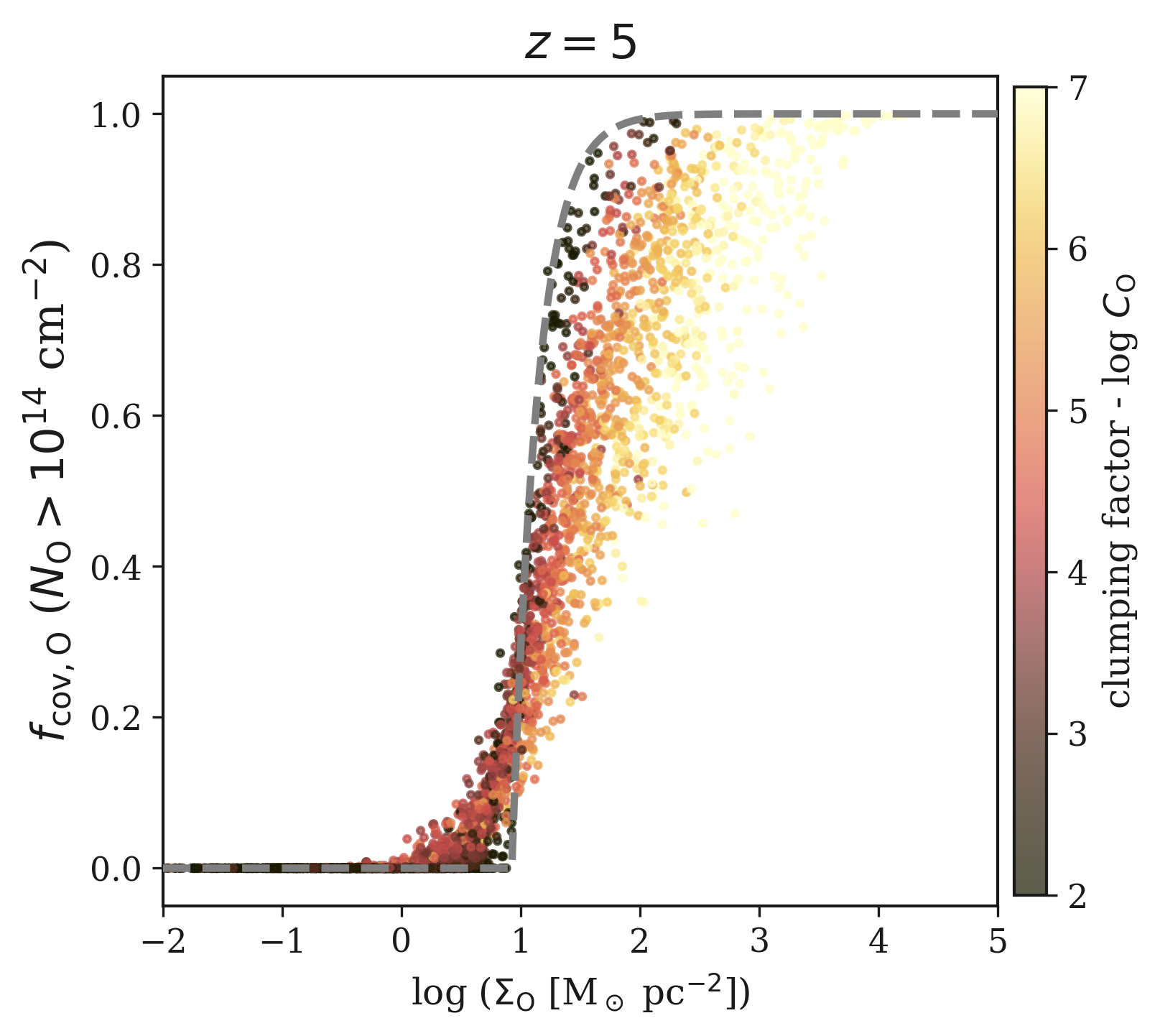}
    \caption{Oxygen covering fractions as a function of oxygen surface density at $z=5$. Points are coloured depending on the oxygen clumping factor, such that dark points have a more homogeneous distribution than lighter ones. The grey line is the expected relation if oxygen was homogeneously distributed inside the {virial radius} of haloes (Eq.~\ref{eq:covf_logsigma}). Haloes with lower oxygen clumping factors follow the expected relation from the homogeneous case more closely.}
    \label{fig:O_clump_exp}
\end{figure}

To test this model, we show a comparison between the theoretical oxygen covering fraction and the measured one, when selecting a threshold of $10^{14}$ cm$^{-2}$, since for this set of data the trend can be better appreciated, without further quantitative analysis.
Data points in Fig.~\ref{fig:O_clump_exp} represent the median \ion{O}{} covering fractions of haloes more massive than $10^8\,\text{M}_\odot$ as a function of their oxygen surface density ($\Sigma_{\rm O}$) at $z=5$.
Each point is coloured according to the value of the oxygen clumping factor inside the CGM of the halo, defined as $C = \left< n_{\rm O}^2 \right>/\left< n_{\rm O} \right>^2$.
Haloes where oxygen is distributed more homogeneously are represented with darker colours, while lighter colours are used to identify haloes with clumpier oxygen distributions.
The dashed gray line defines the relation in Eq.~(\ref{eq:covf_logsigma}).

We note that at very small surface densities, the absence of lines of sight with column density above the threshold of $10^{14}$ cm$^{-2}$ within the virial radius of small haloes, causes their covering fraction to be $0$.
However, close to the turning point, higher clumping factors can help small haloes to reach higher column densities close to the centre, increasing their covering fraction.
At higher surface densities, a higher clumping factor causes a decrease in the covering fraction, concentrating the vast majority of oxygen close to the centre.
Overall, we conclude that haloes with lower clumping factors (darker points), where we expect a more homogeneous distribution of oxygen, seem to closely follow the predicted curve for the covering fraction.
On the other hand, the clumpier the halo is, the further away it is from the theoretical relation.

\section{Best-fit parameters for the \ion{O}{i} cross-section}
\label{app:best-fitsigma}

In Table~\ref{tab:best-fit} we report the best-fit parameters for the broken power law used to modelled $\sigma(M_{\rm DM})$ obtained with a non-linear least squares method.
Each part of the curve follows the relation $\log_{10} (\sigma/{\rm pkpc}^2)=m \log_{10}(M_{\rm DM}/\text{M}_\odot) +q$, with best-fit parameters that shift from $(m_l, q_l)$ to $(m_h, q_h)$ at a value of the dark matter mass $M_c$, which, by imposing continuity, has to be $\log_{10} (M_c/\text{M}_\odot) = (q_h - q_l)/(m_l - m_h)$.

\begin{table}
    \centering
    \begin{tabular}{c||cccccc}
        \hline
          & $z$ & $m_l$ & $q_l$ & $m_h$ & $q_h$ & log$ (M_c/\text{M}_\odot)$ \\
        \hline
        \hline
        \multirow{4}{0.5em}{\rotatebox[origin=c]{90}{$\sigma_{14}$ [pkpc$^2$]}}& 5 & 6.5567 & $-$55.7541 & 0.7110 & $-$4.9131 & 8.6970 \\
        & 6 & 4.7190 & $-$39.5764 & 0.6854 & $-$4.7588 & 8.6319 \\
        & 7 & 3.1988 & $-$26.5364 & 0.6252 & $-$4.2642 & 8.6543 \\
        & 8 & 2.8800 & $-$23.8872 & 0.5351 & $-$3.4997 & 8.6941 \\
        \hline\hline
        \multirow{4}{0.5em}{\rotatebox[origin=c]{90}{$\sigma_{13}$ [pkpc$^2$]}}
        & 5 & 3.3183 & $-$27.4579 & 0.6535 & $-$4.2431 & 8.7116 \\
        & 6 & 2.0184 & $-$16.1433 & 0.6235 & $-$4.0470 & 8.6716 \\
        & 7 & 1.6016 & $-$12.5790 & 0.5737 & $-$3.6674 & 8.6686 \\
        & 8 & 1.4382 & $-$11.2571 & 0.4870 & $-$2.9436 & 8.7400 \\
        \hline\hline
        \multirow{4}{0.5em}{\rotatebox[origin=c]{90}{$\sigma_{12}$ [pkpc$^2$]}}
        & 5 & 4.0306 & $-$33.3923 & 0.6492 & $-$4.1411 & 8.6508 \\
        & 6 & 1.4767 & $-$11.4045 & 0.6183 & $-$3.9468 & 8.6876 \\
        & 7 & 1.1535 & $-$8.6658 & 0.5682 & $-$3.5708 & 8.7040 \\
        & 8 & 1.0292 & $-$7.6771 & 0.4825 & $-$2.8617 & 8.8079 \\
        \hline\hline
    \end{tabular}
    \caption{Best-fit parameters of the broken power law used to model the $\sigma - \log M_{\rm DM}$ relation. Each block represents best-fit values for different column density thresholds ($10^{14}$\,cm$^{-2}$, $10^{13}$\,cm$^{-2}$, $10^{12}$\,cm$^{-2}$, respectively).
    Columns from left to right present values of: redshift, slope and intercept of the curve at $\log M < \log M_c$, slope and intercept of the curve at $\log M > \log M_c$, and finally the value of $\log M_c$ at which the curve shifts from the first power law to the second.}
    \label{tab:best-fit}
\end{table}

%%%%%%%%%%%%%%%%%%%%%%%%%%%%%%%%%%%%%%%%%%%%%%%%%%

% Don't change these lines
\bsp	% typesetting comment
\label{lastpage}
\end{document}